\documentclass[journal]{IEEEtran}
\usepackage{hhline}
\usepackage{xcolor}

\ifCLASSINFOpdf
\else
\fi
\usepackage{amsmath}
\usepackage{amsmath}
\DeclareMathOperator*{\argmax}{arg\,max}
\DeclareMathOperator*{\argmin}{arg\,min}

\usepackage{algorithm}
\usepackage{algpseudocode}
\usepackage{amssymb}
\usepackage{amsmath}
\usepackage{latexsym}
\usepackage{url}
\usepackage{graphicx}

\begin{document}
%
\title{Soft Video Multicasting Using Adaptive Compressed Sensing}
%
%
%

\author{Hadi~Hadizadeh and Ivan V. Baji\'c

\thanks{Hadi Hadizadeh is with the Quchan University of  Technology (e-mail: h.hadizadeh@qiet.ac.ir). Ivan V. Baji\'c is with the School of Engineering Science at Simon Fraser University, Burnaby, BC, V5A 1S6, Canada (e-mail:ibajic@ensc.sfu.ca). The corresponding author is Hadi Hadizadeh.}
\thanks{This work was supported in part by INSF grant number 96010820.}
}

\IEEEpubid{\begin{minipage}{\textwidth}\ \\[12pt]
\\
\\
\\
\\
\copyright 2020 IEEE.  Personal use of this material is permitted. Permission from  
IEEE must be obtained for all other uses,  in any current or future media, 
including reprinting/republishing this material for advertising or promotional
purposes, creating new collective works, for resale or redistribution to servers
or lists, or reuse of any copyrighted component of this work in other works.
\end{minipage}}

\maketitle

\begin{abstract}
Recently, soft video multicasting has gained a lot of attention, especially in broadcast and mobile scenarios where the bit rate supported by the channel may differ across receivers, and may vary quickly over time. Unlike the conventional designs that force the source to use a single bit rate according to the receiver with the worst channel quality, soft video delivery schemes transmit the video such that the video quality at each receiver is commensurate with its specific instantaneous channel quality.
In this paper, we present a soft video multicasting system using an adaptive block-based compressed sensing (BCS) method. The proposed system consists of an encoder, a transmission system, and a decoder. At the encoder side, each block in each frame of the input video is adaptively sampled with a rate that depends on the texture complexity and visual saliency of the block. The obtained BCS samples are then placed into several packets, and the packets are transmitted via a channel-aware OFDM (orthogonal frequency division multiplexing) transmission system with a number of subchannels.
At the decoder side, the received BCS samples are first used to build an initial approximation of the transmitted frame. To further improve the reconstruction quality, an iterative BCS reconstruction algorithm is then proposed that uses an adaptive transform and an adaptive soft-thresholding operator, which exploits the temporal similarity between adjacent frames to achieve better reconstruction quality. The extensive objective and subjective experimental results indicate the superiority of the proposed system over the state-of-the-art soft video multicasting systems.  
\end{abstract}

\begin{IEEEkeywords}
SoftCast, MultiCast, saliency, OFDM, compressed sensing
\end{IEEEkeywords}

%
\IEEEpeerreviewmaketitle

\section{Introduction}
\label{sec:intro}

The traditional image/video communication systems are designed based on the Shannon's source-channel separation theorem, which states that source coding can be separated from channel coding without loss of optimality if the communication channel is \textit{point-to-point}, and if the channel statistics are known to the source \cite{softcast0, softcast1}. However, in practical multicast/broadcast scenarios where the channel is not necessarily point-to-point, and the channel condition and statistics fluctuate in an unpredictable manner, this theorem does not hold anymore. For instance, when broadcasting video to several mobile users, the channel quality of each user may change rapidly due to user movement. Therefore, in such cases, the traditional systems are not very efficient because the transmitted content is usually conservatively encoded at the bit rate supported by the worst receiver to ensure that even the worst receiver is able to achieve an acceptable level of quality \cite{coding_gain}. Using such schemes, users with better channel quality cannot enjoy better quality.

In conventional video communication systems, videos are encoded into a compressed binary stream by quantization and entropy coding followed by channel coding and modulation. Because of extreme compression, the resultant bitstream is very sensitive to bit errors, and even a single bit error may make a complete slice or frame or even several frames useless. Hence, if the channel quality falls below a threshold, the receivers may not be able to decode the video, or the video quality may deteriorate significantly. On the other hand, if the channel quality increases, the video quality at the receivers may not improve accordingly. This phenomenon is known as the \textit{cliff effect} \cite{coding_gain, cliff}.

\subsection{Related Works}
To tackle these challenges, several cross-layer joint source and channel coding approaches have been proposed. For example, a novel  end-to-end approach called SoftCast \cite{softcast0, softcast1} was proposed for scalable wireless video transmission using a pseudo-analog scheme, which has attracted much research attention in recent years \cite{softcast1}, \cite{softcast2}, \cite{softcast3}, \cite{coding_gain}. In contrast to conventional systems, SoftCast unifies both source coding and channel coding into a single framework, in which a video is encoded both for compression and error protection \cite{softcast1}. Unlike the conventional designs, SoftCast does not use quantization and entropy and channel coding. In fact, it only applies a decorrelating linear transform like 3D-DCT (discrete cosine transform) on the video, and produces a stream of real numbers. The resultant numbers are then directly transmitted by raw orthogonal frequency division multiplexing (OFDM) \cite{OFDM} with a dense constellation. Since only linear operations are used in the entire process of SoftCast, the received video quality varies with the channel quality smoothly. Therefore, SoftCast does not suffer from the cliff effect, and each receiver achieves a graceful video quality in a scalable manner according to its channel condition. However, the bit rate produced by SoftCast is relatively high, and also it does not exploit channel information.

In the literature, inspired by SoftCast, several efforts have been made to improve the reconstructed video quality. For instance, in \cite{dcast}, a distributed soft video broadcasting framework called Dcast was proposed, in which the inter-frame redundancy is exploited by a frame prediction scheme to reduce the bit rate of the transmitted video. Also, the predicted frame is utilized as a side information of the distributed video coding at the receiver to improve the reconstructed video quality. To reduce the transmission power
cost, coset codes \cite{coset} are used in Dcast. In \cite{layercast}, another framework called LayerCast was proposed, in which DCT coefficients are processed into multiple layers by coset coding to satisfy users with different channel bandwidth. Specifically, only the base layer is delivered to the users with narrow bandwidth while users with wider bandwidth can use both the base layer and the enhancement layers. Another framework called LineCast was proposed in \cite{linecast} that extends SoftCast for broadcasting the satellite images in real time. In LineCast the satellite images are transmitted line by line in progressive manner, a feature that cannot be achieved by SoftCast. Xiong et al. \cite{gcast} developed a gradient-based image transmission scheme called Gcast which, like SoftCast, transmits an image by sending its gradient data. In \cite{hda}, a hybrid digital-analog (HDA) scalable framework was proposed, in which a base layer is encoded by H.264 \cite{richardson} and an enhancement layer is encoded by SoftCast.

Recently, compressed sensing (CS) \cite{cs, cs2} has drawn great attention in various applications. CS states that if a signal is sparse in a transform domain, it is possible to reconstruct it from far fewer samples than required by the Shannon-Nyquist sampling theorem \cite{cs2}. Several CS-based methods have been developed for wireless video multicasting, which do not suffer from the cliff effect similar to SoftCast. For instance, a distributed compressed sensing (DCS) method was proposed in \cite{dcs_video}, which exploits the temporal correlation of video frames to reduce the bit rate of the transmitted video. In \cite{scale}, a wireless video multicasting scheme was proposed based on multi-scale compressed sensing. A distributed compressed sensing-based multicast method called DCS-cast was proposed in \cite{dcs} that has good performance when the packet loss rate is high. In \cite{ardcs_cast}, an adaptive residual-based DCS scheme called ARDCS-cast was proposed for soft video multicasting. The reported results show that ARDCS-cast outperforms DCS-cast and SoftCast, thanks to the adaptive residual measurement.

The above-mentioned approaches do not use any channel information. However, there are methods that exploit the channel information
to improve the performance of SoftCast. For example, the ParCast method proposed in \cite{parcast}, transmits videos over  multiple-input
multiple-output (MIMO) wireless channels in which the wireless channels are decomposed into subchannels by OFDM. The DCT coefficients are then assigned to subchannels based on the respective sorted order of their energy levels. For example,  high-energy DCT coefficients are transmitted in high-gain channels. Also, ParCast performs joint source-channel power allocation to optimize the total error performance. Later in \cite{parcastplus}, the authors proposed ParCast+ to further improve the performance of ParCast through utilizing motion-compensated temporal filtering (MCTF) to better de-correlate the videos. In \cite{sharpcast}, a hybrid digital
and analog transmission approach called SharpCast was proposed that divides a video into a structure part and a content part. The structure part and the high-energy DCT coefficients of the content part are transmitted digitally, and the rest of the content part is transmitted in analog as SoftCast to reduce the energy consumption. Most of the aforementioned approaches have been developed for real-time applications. In \cite{mcast}, a new linear video transmission scheme called Mcast was proposed, in which the video data is transmitted multiple times across multiple time slots and multiple channels to exploit the time and frequency diversities to improve the received quality. The reported results show that Mcast outperforms SoftCast and ParCast. However, Mcast is not intended for real-time applications.

Another important issue that must be considered in wireless video multicasting is the efficient usage of the transmission power. In SoftCast, in order to utilize the total transmission power efficiently, the encoder scales a group of transform coefficients by a separate scalar within a power-distortion optimization (PDO) framework in which the scaling factor of each group is determined by the expected energy of the coefficients without any perceptual considerations \cite{softcast1}. In the end, coefficients with higher expected energy get more transmission power and vice versa regardless of their perceptual and visual importance. A similar PDO framework is used in almost all the aforementioned approaches. In fact, in all the aforementioned approaches, different parts of the video are treated equally for both power and bit rate allocation.

However, there are often many perceptual redundancies in images/videos \cite{chou, perceptual, watson_book}, and not all parts of an image/video have the same visual importance \cite{ours1}. In fact, it is known that due to the visual attention (VA) mechanism of the human brain \cite{itti_book}, only some of the more interesting regions in the image are attended and perceived consciously by the observer when watching an image. Such regions are usually referred to as \textit{salient regions} \cite{itti_book, IKN}, and their location and visual significance can be described by a \textit{saliency map} \cite{IKN}, which can be estimated by an appropriate computational model of VA \cite{IKN, itti_foveation}. Given that the transmitted videos are ultimately viewed by human observers, it is reasonable to allocate bit rate and power for different regions according to their visual importance so as to use the limited bandwidth and total transmission power more efficiently.
\subsection{Summary of Contributions}
In this paper, we present a DCS-based video multicasting framework that is targeted to meet the following main design goals: 1) The system must be implemented in a distributed manner so that the complexity at the sender side remains as low as possible. 2) The system must use the available bandwidth efficiently by exploiting the visual importance of various blocks in the video. 3) The system must take the channel information into account to achieve better performance, and use the total transmission power efficiently. 4) Similar to SoftCast, the system must deliver the video to the receivers without suffering from the cliff effect.

To achieve the first goal, we use block-based CS (BCS) \cite{fowler_conf, fowler_video} to sample different blocks in an input video at a very low computational complexity, and similar to the distributed systems, we transfer the main complexity to the receivers. In particular, we propose a novel BCS reconstruction method that is used to reconstruct the transmitted frames based on their BCS samples at the receiver side. For the second goal, we propose an adaptive rate control in which the sampling rate of various blocks in the input video is determined based on their visual importance as well as their texture complexity. To achieve the third goal, we first propose an error-resilient packetization scheme to carry the produced BCS samples. We then use an OFDM transmission system with a number of different subchannels to combat with multipath fading. Each produced packet is assigned to a subchannel using a sub-optimal subchannel allocation scheme that uses the available channel information. To use the total transmission power efficiently, we use an optimal power allocation scheme by which a specific power is assigned to each packet on its allocated subchannels according to a packet importance metric. Finally, to meet the fourth goal, we transmit all packets in a pseudo-analog manner similar to SoftCast without using any non-linear operation like quantization and entropy coding.

The main contributions of this paper are as follows: 1) We propose an adaptive rate control for soft video multicasting by which the bit rate produced by the proposed system can easily be controlled at the block level based on a specific target bit rate. In particular, the proposed rate control algorithm considers  the effect of visual attention, so it can be considered as a perceptual rate control. None of the aforementioned works has this feature. 2) We propose a BCS reconstruction algorithm in which we develop an adaptive transform, and an adaptive soft-thresholding operator to improve sparsity, thereby improving the reconstruction quality.

To evaluate the proposed system, several objective and subjective experiments were conducted. The results indicate that the proposed system outperforms various existing approaches including SoftCast \cite{softcast1}, DCS-cast \cite{dcs}, ARDCS-cast \cite{ardcs_cast}, and ParCast+ \cite{parcastplus}.

The paper is organized as follows. In Section \ref{sec:BCS}, we briefly review some background information about BCS. The proposed system is then presented in Section \ref{sec:proposed}. The experimental results are given in Section \ref{sec:results} followed by conclusions in Section \ref{sec:conclusions}. In this paper, capital bold letters (e.g., $\mathbf{X}$) denote matrices, lowercase bold letters (e.g., $\mathbf{x}$) denote vectors, and italic letters (e.g., $t$ or $T$) represent scalars.

\section{Background}
\label{sec:BCS}

In this section, we briefly describe the BCS process. Let $\mathbf{X}$ be an input image. To apply BCS, the image is first divided into small non-overlapping blocks of size $B\times B$, and each block is sampled separately. Let $\mathbf{x}_j$ be the vectorized signal of the $j$-th block, which is obtained by concatenating all columns of $\mathbf{X}$ to each other. The corresponding output CS vector, $\mathbf{y}_j$, (of length $C$) is obtained as:
\begin{equation}
\label{eq:bcs}
\mathbf{y}_j = \pmb \Phi_B \mathbf{x}_j,
\end{equation}
where $\pmb \Phi_B \in \mathbb{R}^{C\times B^2}$ is the sampling or measurement matrix, which is an orthonormalized i.i.d Gaussian matrix \cite{cs2}. In fact, (\ref{eq:bcs}) provides a simple structure at the sender side for signal sampling with a very low computational complexity. At the receiver side, if $\pmb\Phi_B$ is a full-rank matrix, an initial approximation of $\mathbf{x}_j$ can be obtained by $\hat{\mathbf{x}}_j = \pmb\Phi_B^{\dag} \mathbf{y}_j$, where the superscript $\dag$ denotes the pseudo inverse. Since usually $C\ll B^2$, recovering every $\mathbf{x}_j \in \mathbb{R}^{B^2}$ from its corresponding $\mathbf{y}_j$ is impossible in general. However, CS theory states that if $\mathbf{x}_j$ is sufficiently sparse in a transform ($\pmb\Psi$) domain, exact recovery is possible. Several recovery methods have been developed for this purpose \cite{J1,J2,J3,J4}. Specifically, if the transform coefficients, $\mathbf{v}_j=\pmb\Psi \mathbf{x}_j$, are sufficiently sparse, the solution of the recovery procedure can be found with several $l_0$ optimization procedures or their $l_1$-based convex relaxations that use pursuit-based methods \cite{cs, cs2}. However, the computational complexity of such methods is often too high, especially for images and videos.

To reduce the computational complexity, several BCS reconstruction methods have been proposed \cite{fowler_conf, bcs_spl, fowler_video}. The popular ones are the iterative-thresholding reconstruction algorithms that start from some initial approximation and form the approximation at each iteration using a specific instance of a projected Landweber (PL) algorithm \cite{pl}. Among such PL-based algorithms is the popular BCS-SPL algorithm \cite{bcs_spl} that incorporates a smoothing operation (like Wiener filtering) at each iteration to reduce blocking artifacts. This imposes smoothness in addition to the sparsity inherent to PL. The BCS-SPL algorithm is described in Algorithm \ref{alg:bcs} \cite{bcs_spl}. As seen from this algorithm, at each iteration $k$, the current reconstruction $\mathbf{X}^k$  is first smoothed with a Wiener filter to obtain $\hat{\mathbf{X}}^k$ . Then each block in $\hat{\mathbf{X}}^k$, i.e. $\hat{\mathbf{x}}_j^k$ is projected onto a convex set (hyper-plane) $\mathcal{C}=\{\mathbf{g}:\mathbf{y}=\pmb\Phi \mathbf{g}\}$ to obtain $\tilde{\mathbf{x}}_j^k$  as follows:
\begin{equation}
\tilde{\mathbf{x}}_j^k = \hat{\mathbf{x}}_j^k + \pmb\Phi_B^t(\mathbf{y}_j-\pmb\Phi_B\hat{\mathbf{x}}_j^k),
\end{equation}
where $\pmb\Phi_B^t$ is the transpose of $\pmb\Phi_B$. In fact,  $\tilde{\mathbf{x}}_j^k$ is the closest vector to $\hat{\mathbf{x}}_j^k$ on $\mathcal{C}$.
 The transform coefficients of the image $\tilde{\mathbf{X}}^k$ are then obtained as $\hat{\mathbf{v}}^k = \pmb\Psi \tilde{\mathbf{x}}^k$, where $\pmb\Psi$ is a fixed transform like DCT,  contourlets or complex-valued wavelets. After that, a hard-thresholding operator $\mathcal{H}(\cdot)$ is applied on $\hat{\mathbf{v}}^k$ to reduce Gaussian noise. The inverse transform, $\pmb\Psi^{-1} $, is then applied on the noise-reduced transform coefficients  to yield the reconstructed image  $\bar{\mathbf{X}}^k$. Finally, $\bar{\mathbf{X}}^k$  is again projected back to the convex set $\mathcal{C}$ to obtain $\mathbf{X}^k$. This procedure is repeated several times until a convergence or stopping criteria are met. The $\mathbf{X}^k$ at the last iteration will then be returned as the best approximation of the transmitted image $\mathbf{X}$.

BCS-SPL is a flexible algorithm because it makes it possible to incorporate sophisticated transforms and thresholding operators, as well as additional constraints into its iterative procedure \cite{bcs_spl, fowler_video}. Hence, we adopt the BCS-SPL framework as the core frame reconstruction algorithm in this paper, but with some modifications.

\begin{algorithm}
\caption{The original BCS-SPL algorithm \cite{fowler_conf}.}
\label{alg:bcs}
\begin{algorithmic}[1]
\Procedure{$\mathbf{X}=$BCS\_SPL}{$\mathbf{y},\pmb\Phi_B, \pmb\Psi,k_{\max}, \epsilon$}
\State \textbf{Initialize:} Set $k=0$.
\State For each block $j$ do $\mathbf{x}_j^k=\pmb\Phi_B^t\mathbf{y}_j$ 
    \Repeat
\State $\hat{\mathbf{X}}^k \gets$ SmoothingFilter($\mathbf{X}^k$) 
      \For{each block $j$}
        \State $\tilde{\mathbf{x}}_j^k \gets \hat{\mathbf{x}}_j^k + \pmb\Phi_B^t(\mathbf{y}_j-\pmb\Phi_B\hat{\mathbf{x}}_j^k)$
      \EndFor
	\State $\hat{\mathbf{v}}^k \gets \pmb\Psi \tilde{\mathbf{x}}^k$
	\State $\mathbf{v}^k \gets \mathcal{H}(\hat{\mathbf{v}}^k)$
	\State $\bar{\mathbf{x}}^k \gets \pmb\Psi^{-1} \mathbf{v}^k$
      \For{each block $j$}
	\State $\mathbf{x}_j^{k+1} \gets \bar{\mathbf{x}}_j^k + \pmb\Phi_B^t(\mathbf{y}_j-\pmb\Phi_B\bar{\mathbf{x}}_j^k)$
	\EndFor
	\State Compute error $e^{k+1} \gets \|\mathbf{X}^{k+1} - \tilde{\mathbf{X}}^{k}\|_2$
	\State $k \gets k+1$
	\Until{$|e^{k}-e^{k-1}|<\epsilon$ or $k=k_{\max}$}
	\State $\mathbf{X} \gets \mathbf{X}^{k}$
\EndProcedure \\

Note: $k_{\max}$ is the maximum number of iterations, and $\epsilon$ denotes the tolerance.
$\mathbf{x}_j^k$ denotes the $j$-th block of frame $\mathbf{X}$ in the $k$-th iteration. $\mathcal{H}(\cdot)$ is the hard-thresholding operator.
\end{algorithmic}
\end{algorithm}

\section{The Proposed Method}
\label{sec:proposed}
\begin{figure}
\centering
\includegraphics[scale=0.42]{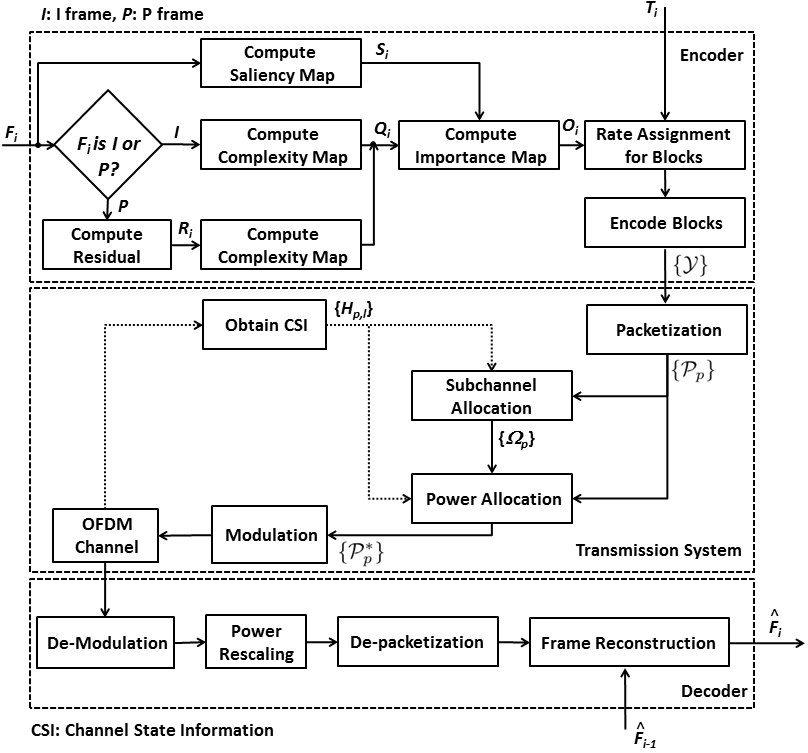}
\caption{The flowchart of the proposed method for encoding, transmitting and recovering a single frame $\mathbf{F}_i$. Here, $\{\mathcal{Y}\}$ denotes the set of the BCS samples of all blocks of frame $\mathbf{F}_i$.  Similarly, $\{\mathcal{P}_p\}$ and $\{\mathcal{P}_p^{*}\}$ denote the set of all packets after packetization and power allocation, respectively. Also, $\{\Omega_p\}$ denotes the set of all subchannel allocation sets related to all the $P$ packets. $\hat{\mathbf{F}}_i$ is the reconstructed version of $\mathbf{F}_i$. $T_i$ is the target rate for $\mathbf{F}_i$.}
\label{fig:flowchart}
\end{figure}

In this section, we present the proposed soft video multicasting system. The proposed system consists of an encoder, a transmission system, and a decoder. The flowchart of the proposed system is depicted in Fig. \ref{fig:flowchart}. In the sequel, we describe the details of each part.
\subsection{The proposed video encoder}
\label{sec:encoder}
Let $\mathcal{V}$ be an input video sequence consisting of $N$ video frames, i.e. $\mathcal{V}=\{\mathbf{F}_i\}$ for $i=1,\cdots,N$. Assume that each video frame is of resolution $W\times H$. The video sequence is divided into a number of GOPs (groups of pictures) of length $L_{GOP}$, where the first frame in each GOP is an I frame, and the remaining frames are P frames. In the proposed method, each I frame is encoded by the proposed frame encoding method directly and independently while each P frame is first predicted from its previous frame to obtain a residual frame, and the residual frame is then encoded by the proposed frame encoding method. Specifically, to maintain the encoding complexity as low as possible, the residual frame ($\mathbf{R}_i$) for the $i$-th P frame is computed by a simple frame subtraction as 
$\mathbf{R}_i = \mathbf{F}_{i} - \mathbf{F}_{i-1},$ for $i\geq 2$. Note that, since $\mathbf{R}_i$ may be sparser than $\mathbf{F}_i$, it can be reconstructed better than $\mathbf{F}_i$ when using a suitable CS reconstruction method \cite{ardcs_cast}. In fact, due to the similarity of  adjacent frames, the residual frames are usually very sparse. Hence, fewer CS measurements can be used to encode the residual frames so as to use the available bandwidth more efficiently. 

\subsubsection{The proposed frame encoding method}
In the proposed frame encoding method, each frame (an I or a residual frame) is divided into $M$ non-overlapping blocks of size $B\times B$ pixels, where $M=W\times H/B^2 $. Here, we assume that $M$ is always an integer. Each block is vectorized to obtain a column vector by concatenating all columns to each other. Let $\mathbf{x}_j \in \mathbb{R}^{B^2}$ be the vectorized version of the $j$-th block, where $j=1,\cdots, M$.
We then use BCS to obtain the CS measurements of each block independently. Specifically, the CS measurements for the $j$-th block, $\mathbf{y}_j$, are obtained as $\mathbf{y}_j = \pmb\Phi_{j} \mathbf{x}_j,$ where $\pmb\Phi_j \in \mathbb{R}^{m_j\times B^2}$ denotes the block sampling matrix, where $m_j$ is the length of $\mathbf{y}_j$. If $m_j=B^2$, the sampling is performed at full rate, and when $m_j<B^2$, the sampling is performed at a lower rate. The sampling rate for the $j$-th block is defined as $\rho_j=m_j/B^2$. In practice, we need at least a few measurements ($m_{min}$) so as to be able to reconstruct a block with a reasonable quality. Hence, $m_{min}\leq m_j\leq B^2$, so $\rho_{min}\leq \rho_j \leq 1$, where $\rho_{min} = m_{min}/B^2$. In this paper, we use $8\times 8$ blocks, hence $B^2=64$. In Section \ref{sec:results}, we analyze the performance of the proposed system with other block sizes as well. Also, we use $m_{min}=10$ and $\rho_{min}=0.15625$.

In conventional BCS schemes, the same block sampling matrix is used for all blocks, so all blocks are sampled with the same sampling rate. However, in this paper, we present an adaptive scheme by which the sampling rate of various blocks varies according to their importance and the available rate. For this purpose, we first propose a scheme in Section \ref{sec:frame_rate} to compute a total (target) rate for each video frame. We then propose an adaptive scheme in Section \ref{sec:block_rate} to compute the sampling rate of each block.

\subsubsection{Rate allocation for each video frame}
\label{sec:frame_rate}
Given that different frames have different complexity, they may require different rates for encoding. In this section, we propose a scheme by which the total rate of each I or residual frame is computed based on its normalized complexity. 

Let $T_{tot}$ be the total rate budget of the input video sequence $\mathcal{V}$, which is given by the user or inferred by monitoring the channel. In other words, $T_{tot}$ CS samples (measurements) must be produced for the entire video, among which $T_h$ samples are used for overhead and metadata, and $T_d=T_{tot}-T_h$ samples are used for actual data. Also, let $T_{min}$ denote the minimum number of the required CS samples for each frame. In fact, $T_{min}=M\cdots m_{min}$. Since different frames have different complexity, we compute the rate budget for frame $\mathbf{F}_i$, i.e. $T_i$ , based on its complexity, $C_i$, so that frames with higher complexity get higher rate and vice versa. For this purpose, we use the following approach. We first allocate $m_{min}$ samples to all blocks in each frame. In other words, we first allocate $T_{min}$ samples to each frame. We then assign $T_i^a$ additional samples to the $i$-th frame as follows: 
\begin{equation}
\label{eq:Ti}
T_i^a=\text{round}\Big(\frac{C_i}{\sum_{k=1}^N C_{k}}\cdot (T_{d} - N\cdot T_{min})\Big),
\end{equation}
where $i=1,\cdots, N$. Note that $C_i/\sum_{k=1}^N C_{k}$ can be considered as the normalized complexity of the $i$-th frame. Hence, using (\ref{eq:Ti}), the value of $T_i^a$ for the $i$-th frame is proportional to the frame normalized complexity. Up to this point, the rate assigned for each frame is equal to $T_{min} + T_i^a$. Because of the rounding process in (\ref{eq:Ti}),  the sum of the assigned rates for all frames may not equal $T_d$, and an overflow or underflow $\zeta$ may happen, where its value can be computed as follows:
\begin{equation}
\zeta = T_d - \sum_{i=1}^N (T_{min} + T_i^a),
\end{equation} 
where $\zeta>0$ indicates an underflow and $\zeta<0$ indicates an overflow. To meet the target data rate $T_d$ in the case of an underflow, we assign $\zeta$ samples equally to the $\zeta$ frames with the highest normalized complexity, and in the case of an overflow, we remove $\zeta$ samples equally from the $\zeta$ frames with the lowest normalized complexity. This way, we can make sure that $\sum_{i=1}^N T_i =T_d$. As will be discussed later, a few metadata is sent for each frame, where the sum of all metadata for the entire video is equal to $T_h$.

To compute frame complexity, different methods can be used. Here, to keep the computational complexity low, we propose to use the total variation (TV) of each frame as its complexity, which can be computed for $\mathbf{F}_i$ as follows:
\begin{equation}
C_i= \frac{1}{W\cdot H} \sum_{y=1}^H \sum_{x=1}^W G_i(x,y),
\end{equation}
where $G_i(x,y)$ is the gradient magnitude at location $(x,y)$ of $\mathbf{F}_i$. In this paper, we use the Sobel operators \cite{gonzalez} to compute the gradient magnitude of each frame. 

\subsubsection{Rate allocation for each block}
\label{sec:block_rate}
After computing $T_i$ for frame $\mathbf{F}_i$, we propose to allocate $T_i$ samples to the blocks in $\mathbf{F}_i$ adaptively based on their importance. Specifically, we first allocate $m_{min}$ samples to each block in the frame. We then allocate $m_j^a$ additional samples to each block as follows: 
\begin{equation}
\label{eq:ri}
m_j^a=\text{round}\Big(\frac{o_j}{\sum_{m=1}^M o_m}\cdot (T_i-M\cdot m_{min})\Big),
\end{equation}
where $j=1,\cdots,M$, and $o_j$ is the importance value of the $j$-th block. Note that $o_j/\sum_{m=1}^M o_{m}$ can be considered as the normalized importance of the $j$-th block. Hence, using (\ref{eq:ri}), the value of $m_j^a$ for each block is proportional to the block normalized importance. Because of the rounding process in (\ref{eq:ri}), the sum of the assigned rates for all blocks may not equal $T_i$, and an overflow or underflow $\kappa$ may happen, whose value can be computed as follows:
\begin{equation}
\kappa = T_i - \sum_{i=1}^M (m_{min} + m_j^a),
\end{equation} 
where $\kappa>0$ indicates an underflow and $\kappa<0$ indicates an overflow. To meet the target rate $T_i$ in the case of an underflow, we assign $\kappa$ samples equally to the $\kappa$ blocks with the highest normalized importance, and in the case of an overflow, we remove $\kappa$ samples equally from the $\kappa$ blocks with the lowest normalized importance. This way, we can make sure that $\sum_{j=1}^M m_j =T_i$.
Note that after computing $m_j$ for each block, the sampling matrix $\pmb\Phi_j$ for each block is obtained by simply taking the first $m_j$ rows of a pre-computed general sampling matrix $\pmb\Phi_B \in \mathbb{R}^{B^2\times B^2}$, which is the same for all blocks.

\begin{figure}
\centering
\includegraphics[scale=0.15]{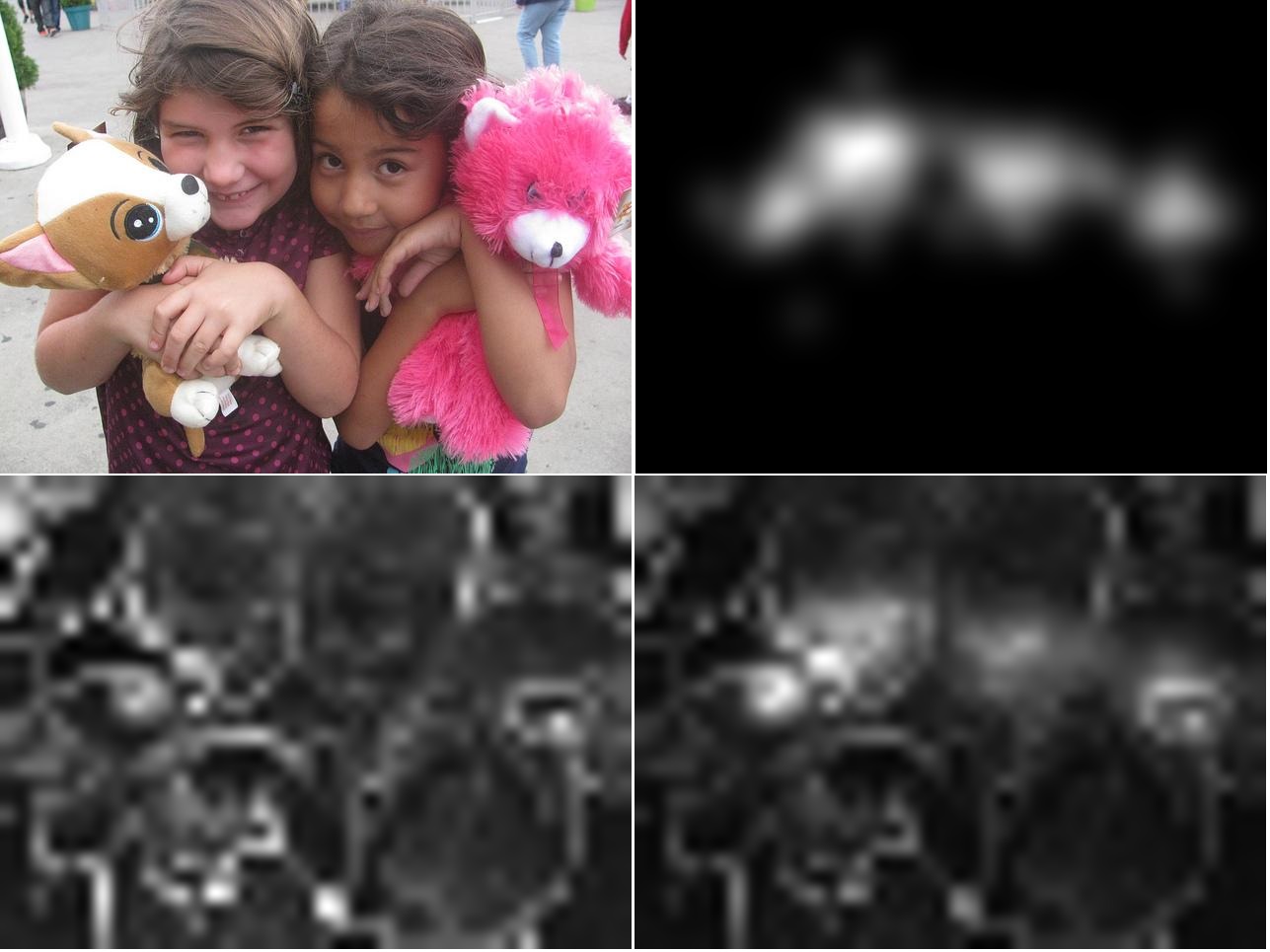}
\caption{A sample image \cite{mit} with its saliency map $\mathbf{S}_i$ (top right), its texture complexity map $\mathbf{Q}_i$ (bottom left), and its importance map $\mathbf{O}_i$ (bottom right).}
\label{fig:importance}
\end{figure}

To compute $o_j$, we use both the texture complexity and visual saliency \cite{itti_book} of the block. To compute the texture complexity of the $j$-th block, $q_j$, we simply use the mean total variation (TV) of the block. In general, blocks with complex texture have higher TV and vice versa. Let $\mathbf{Q}_i$ be the normalized TV map (matrix) of $\mathbf{F}_i$ computed as follows:
\begin{equation}
\mathbf{Q}_i=\Big[\frac{q_j}{q_{\max}}:j=1,\cdots,M\Big],
\end{equation}
where $q_{\max}=\max\{q_j: j=1,\cdots, M\}$. Also, let $s_j$ be the visual saliency of the $j$-th block. Similar to $\mathbf{Q}_i$, we compute a normalized saliency map (matrix), $\mathbf{S}_i$, as follows: 
\begin{equation}
\mathbf{S}_i=\Big[\frac{s_j}{s_{\max}}:j=1,\cdots,M\Big],
\end{equation}
where $s_{\max}=\max\{s_j: j=1,\cdots, M\}$. After computing $\mathbf{Q}_i$ and $\mathbf{S}_i$, we fuse them together to obtain an importance map (matrix), $\mathbf{O}_i$, as follows:
\begin{equation}
\label{eq:pi}
\mathbf{O}_i=\alpha \mathbf{Q}_i + \beta \mathbf{S}_i + \gamma \mathbf{Q}_i \odot  \mathbf{S}_i,
\end{equation}
where $\odot$ represents pixel-wise (Hadamard) multiplication, and $\alpha, \beta$ and $\gamma$ are three constant parameters by which the importance (weight) of each component in (\ref{eq:pi}) can be controlled. In our experiments we found that using $\alpha=\beta=\gamma=1$ is sufficient to achieve good results. The third component in (\ref{eq:pi}) considers the mutual interaction between $\mathbf{Q}_i$ and $\mathbf{S}_i$  \cite{fusion}. In fact, when the value of both $\mathbf{Q}_i $ and $\mathbf{S}_i$ at a location is high, the third term promotes this coherency in $\mathbf{O}_i$ and vice versa. Note that $\mathbf{O}_i=[o_j:j=1,\cdots,M]$. Hence, after computing $\mathbf{O}_i$, the value of its elements can be used in (\ref{eq:ri}) to obtain the sampling rate of each block based on the block importance. This way, we observe that blocks with higher importance get a larger sampling rate and vice versa. Fig. \ref{fig:importance} shows an example image with its corresponding texture complexity, visual saliency, and importance maps.

To compute the visual saliency map of $\mathbf{F}_i$, any visual saliency model like \cite{IKN} can be used. In our expeirments, we used the saliency model proposed in \cite{SR}, whose computational complexity is very low and its accuracy is acceptable.

\subsection{Transmission through the communication channel}
\label{sec:channel}
After adaptive sampling of all blocks in an I or a residual frame, the generated samples are placed in a number of packets, and the packets are then transmitted via the communication channel. Here, we describe the proposed packetization scheme followed by the transmission system.
\subsubsection{The proposed packetization scheme}
In the proposed packetization scheme, CS samples of various blocks are dispersed over different packets to increase the error resilience of the samples. Specifically, we first create $P=B^2$ empty packets, $\{\mathcal{P}_p\}$ for $p=1,2,\cdots,P$. We then put the $p$-th CS sample of all the $M$ blocks of the frame into the $p$-th packet. The blocks are selected in a raster scan order. Given that the number of CS samples in different blocks may be different, as $p$ increases, the length of the packets may decrease because the number of samples of some blocks may be smaller than $p$. Hence, in such cases, some packets may not have any samples of some blocks with lower importance. 
Using this scheme, when a packet gets lost or damaged, only one CS sample in each block is affected. Therefore, the resultant error is dispersed smoothly across the entire frame, not heavily on a single or multiple adjacent blocks. The number of samples in each block is also transmitted to the receivers as metadata so that the receivers can extract the samples of each block from the received packets.

\subsubsection{The transmission system}
Using the above-mentioned packetization procedure, we end up with $P$ packets per each frame. To transmit these packets to the receivers, we use an OFDM system \cite{OFDM} with $L$ subchannels, and with total transmit power constraint $\Gamma_{tot}$. We assume that perfect instantaneous channel state information (CSI) is available at the transmitter side for each channel. We assign each packet to one or a few subchannels, and also assign a specific power to each packet in each of its allocated subchannels. For this purpose, we optimize the subchannel and power allocation to achieve the highest sum error-free capacity under the total transmit power constraint. Specifically, similar to \cite{ofdm_shen}, we use the equally-weighted sum capacity as the objective function with a set of non-linear constraints to impose proportional fairness across different packets to control the capacity ratios among different packets according to their importance. Mathematically, we define the following optimization problem to achieve our goal:
\begin{multline}
\label{eq:opt0}
(g_{p,l}^{*},\zeta^{*}_{p,l}) = \argmax_{g_{p,l},\zeta_{p,l}} \sum_{p=1}^P \sum_{l=1}^L \frac{\zeta_{p,l}}{L} \log_2\Bigg(1+\frac{g_{p,l}h_{p,l}^2}{N_0 \frac{\Theta}{L}}\Bigg), \\
\text{subject to}~\sum_{p=1}^P \sum_{l=1}^L g_{p,l} \leq \Gamma_{tot}, \\
g_{p,l} \geq 0 ~\text{for all}~ p,l \\
\zeta_{p,l}=\{0,1\} ~\text{for all}~ p,l \\
\sum_{l=1}^L \zeta_{p,l}=1 ~ \text{for all} ~n \\
R_1:R_2:\cdots:R_P = \eta_1:\eta_2:\cdots\eta_P
\end{multline}
where $N_0$ is the power spectral density of AWGN (additive white Gaussian noise), and $\Theta$ is the total available bandwidth. Also, $g_{p,l}$ is the power allocated for the $p$-th packet in subchannel $l$, and $h_{p,l}$ is the channel gain for packet $p$ in subchannel $l$. $\zeta_{p,l}$ can only be either 0 or 1, indicating whether subchannel $l$ is used by packet $p$ or not. The fourth constraint shows that only one subchannel can be allocated to each packet. The last constraint imposes proportional fairness among different packets in which $\{\eta_p\}_{p=1}^P$ is a set of constant values that are used to control the capacity ratio of different packets, while $R_p$ denotes the capacity of the $p$-th packet defined as:
\begin{equation}
\label{eq:cap}
R_p = \sum_{l=1}^L \frac{\zeta_{p,l}}{L}\log_2\Big(1+\frac{g_{p,l}h^2_{p,l}}{N_0\frac{\Theta}{L}}\Big).
\end{equation}
Hence, by solving (\ref{eq:opt0}), the sum capacity of all packets are maximized under the total transmit power $\Gamma_{tot}$. This goal is achieved when both the subchannel and power allocation are performed jointly. However, such a joint optimization has a very high computational complexity, which limits the practical application of the proposed method. To mitigate this problem, an alternative suboptimal approach was proposed in \cite{rhee} in which the subchannel and power allocation are peformed separately, and the constraint that subchannels can only be used by one packet is relaxed to make the problem easier. In fact, in the alternative approach, a packet can be assigned to more than one subchannel. However, in this paper, we assume that $L=P$. Hence, each packet is assigned to only one subchannel.
We use this approach for subchannel and power allocation for each packet. For this purpose, we first describe the subchannel allocation procedure.
\subsubsection{Suboptimal subchannel allocation for each packet}
In the suboptimal subchannel allocation algorithm proposed in \cite{rhee}, equal power distribution is assumed accross all subchannels. Let $H_{p,l}=h^2_{p,l}/(N_0 \Theta/L)$ be the channel-to-noise ratio (CNR) for packet $p$ in subchannel $l$, and $\Omega_p$ be the set of subchannels assigned to packet $p$. The allocation procedure is shown in Algorithm \ref{alg:channel}. The main idea of this algorithm is for each packet to use the subchannels with high channel-to-noise ratio as much as possible. At each iteration, the packet with the lowest proportional capacity is able to choose which subchannel to use. Using this approach, high capacity is achieved for all packets, even those with poor channel gains with a low computational complexity. At the end of this procedure, we obtain $\Omega_p$ for each packet $p$, where $p=1,\cdots,P$.
\subsubsection{Optimal power allocation for each packet}
After performing the subchannel allocation procedure, we need to find the optimal power allocation for each packet on its allocated subchannels. In other words, we need to compute $g_{p,l}$ (for all $l \in \Omega_p$) for the $p$-th packet on subchannels determined by $\Omega_p$. We formulate the problem as follows \cite{ofdm_shen}:
\begin{multline}
\label{eq:opt1}
g_{p,l}^{*} = \argmax_{g_{p,l}} \sum_{p=1}^P \sum_{l\in \Omega_p} \frac{1}{L} \log_2\Bigg(1+\frac{g_{p,l}h_{p,l}^2}{N_0 \frac{\Theta}{L}}\Bigg), \\
\text{subject to}~\sum_{p=1}^P \sum_{l\in \Omega_p} g_{p,l} \leq \Gamma_{tot}, \\
g_{p,l} \geq 0 ~\text{for all}~ p,l \\
\Omega_{p} ~\text{are disjoint for all}~p \\
\Omega_{1} \cup \Omega_{2} \cdots \Omega_{P} \subseteq \{1,2,\cdots,L\} \\
R_1:R_2:\cdots:R_P = \eta_1:\eta_2:\cdots\eta_P
\end{multline}
where $\Omega_p$ is the set of all subchannels assigned for packet $p$, and $\Omega_p$ and $\Omega_l$ are disjoint when $p \neq l$. For setting the value of $\eta_p$, we define a packet importance index, $\pi_p$, defined as
$\pi_p = \frac{\mathcal{L}(p)}{\sum_{q=1}^P\mathcal{L}(q)}, $
where $\mathcal{L}(p)$ is the length of the $p$-th packet, i.e. the number of samples in it. The idea is that larger packets carry samples from a larger number of blocks as compared to smaller packets, so they have more importance. We set $\eta_p=\pi_p$ so that packets with higher importance achieve higher capacity as compared to other packets with lower importance.

The optimization problem in (\ref{eq:opt1}) can be solved by finding the maximum of the following cost function:
\begin{multline}
\label{eq:opt2}
J = \sum_{p=1}^P \sum_{l\in \Omega_p} \frac{1}{L} \log_2\Bigg(1+g_{p,l}H_{p,l}\Bigg)  \\
+ \lambda_1 \Bigg( \sum_{p=1}^P \sum_{l\in \Omega_p} g_{p,l}-\Gamma_{tot}\Bigg)  \\
+ \sum_{p=2}^P \lambda_p \Bigg(\sum_{l\in \Omega_1} \frac{1}{L} \log_2(1+g_{1,l}H_{1,l})  \\
- \frac{\eta_1}{\eta_p} \sum_{l\in \Omega_p} \frac{1}{L} \log_2(1+g_{p,l}H_{p,l}) \Bigg), 
\end{multline}
where $\{\lambda_p\}_{p=1}^P$ are the Lagrangian multipliers. To find the maximum of (\ref{eq:opt2}), we differentiate it with respect to $g_{p,l}$, and set each derivative to zero as follows:
\begin{multline}
\label{eq:diff1}
\frac{\partial J}{\partial g_{1,l}}=\frac{1}{L \ln 2} \cdot \frac{H_{1,l}}{1+H_{1,l}g_{1,l}} + \lambda_1 \\
+ \sum_{p=2}^P \lambda_p \frac{1}{L \ln 2} \cdot \frac{H_{1,l}}{1+H_{1,l}g_{1,l}} =0.
\end{multline}
Also, for $p=2,3,\cdots,P$, and $l \in \Omega_p$, we obtain:
\begin{multline}
\label{eq:diff2}
\frac{\partial J}{\partial g_{p,l}}=\frac{1}{L \ln 2} \cdot \frac{H_{p,l}}{1+H_{p,l}g_{p,l}} + \lambda_1 \\
- \lambda_p \frac{\eta_1}{\eta_p} \cdot \frac{1}{L \ln 2} \cdot \frac{H_{p,l}}{1+H_{p,l}g_{p,l}} =0.
\end{multline}
From (\ref{eq:diff1}) and (\ref{eq:diff2}), we can obtain:
\begin{equation}
\label{eq:eq}
\frac{H_{p,m}}{1+H_{p,m}g_{p,m}} = \frac{H_{p,n}}{1+H_{p,n}g_{p,n}},
\end{equation}
for $m,n \in \Omega_p$ and $p=1,\cdots,P$. Without loss of generality, we assume that $H_{p,1} \leq H_{p,2}\leq \cdots \leq H_{p,N_p}$, where $N_p$ is the number of subchannels in $\Omega_p$. Hence, (\ref{eq:eq}) can be written as:
\begin{equation}
\label{eq:gpl}
g_{p,l} = g_{p,1} + \frac{H_{p,l} - H_{p,1}}{H_{p,l}H_{p,1}},
\end{equation}
for $l=1,\cdots,N_p$ and $p=1,\cdots,P$. The above equation shows the power allocation for packet $k$ on subchannel $l$.

Let $\Gamma_{p,tot}$ be the total power allocated for the $p$-th packet, which, using (\ref{eq:gpl}), becomes:
\begin{equation}
\label{eq:ptot}
\Gamma_{p,tot} = \sum_{l=1}^{N_p} g_{p,l} = N_p g_{p,1} + \sum_{l=2}^{N_p} \frac{H_{p,l} - H_{p,1}}{H_{p,l}H_{p,1}},
\end{equation}
for $p=1,2,\cdots, P$. Using (\ref{eq:eq}) and (\ref{eq:ptot}), the capacity ratio constraints in (\ref{eq:opt1}) can be written as:
\begin{multline}
\label{eq:cons1}
\frac{1}{\eta_1} \frac{N_1}{L} \Bigg(\log_2\Big(1+H_{1,1} \frac{\Gamma_{1,tot}-\mu_1}{N_1}\Big)+\log_2 \xi_1 \Bigg) \\
=\frac{1}{\eta_p} \frac{N_p}{L} \Bigg(\log_2\Big(1+H_{p,1} \frac{\Gamma_{p,tot}-\mu_p}{N_p}\Big)+\log_2 \xi_p \Bigg),
\end{multline}
for $p=2,3,\cdots,P$, where $\mu_p$ and $\xi_p$ are defined as:
\begin{equation}
\mu_p = \sum_{l=2}^{N_p}\frac{H_{p,l} - H_{p,1}}{H_{p,l}H_{p,1}},
\end{equation}
and
\begin{equation}
\xi_p = \Bigg(\prod_{l=2}^{N_p}\frac{H_{p,l}}{H_{p,1}}\Bigg)^{\frac{1}{N_p}},
\end{equation}
for $p=1,2,3,\cdots,P$. In addition to (\ref{eq:cons1}), we also have to take the following total power constraints into account:
\begin{equation}
\label{eq:cons2}
\sum_{p=1}^P \Gamma_{p,tot} = \Gamma_{tot}.
\end{equation}
Note that there are $P$ unknowns $\{\Gamma_{p,tot}\}_{p=1}^P$ in the set of $P$ equations in (\ref{eq:cons1}) and (\ref{eq:cons2}). Hence, by solving these $P$ non-linear equations using iterative methods like the Newton-Raphson or quasi-Newton methods \cite{numerical}, we can obtain $\{\Gamma_{p,tot}\}_{p=1}^P$. After that, using (\ref{eq:ptot}) and (\ref{eq:gpl}), the value of $g_{p,l}$ for each packet ($p$) on each subchannel ($l$) can be found. All samples of the $p$-th packet, i.e. $\mathcal{P}_p$, are then scaled by $g_{p,l}$ to obtain packet $\mathcal{P}_p^{*}$, and the scaled samples in $\mathcal{P}_p^{*}$ are transmitted via the related subchannel(s) after modulation with a suitable modulation scheme like 64QAM \cite{dig_comm}. The value of $g_{p,l}$ is also transmitted as metadata so that the receivers can perform the inverse scaling of the received samples.

\begin{algorithm}
\caption{The suboptimal sunchannel allocation algorithm.}
\label{alg:channel}
\begin{algorithmic}[1]
\Procedure{$\{\Omega_p\}_{p=1}^P=$channel\_alloc}{$\{H_{p,l}\}_{p=1,l=1}^{P,L}$}
\State \textbf{Initialize:} Set $R_p=0$, $\Omega_p=\oslash$ for $p=1,2,\cdots,P$ and $\mathcal{A}=\{1,2,\cdots,P\}$.
      \For {$p=1$ to $P$}
      \State $l^{*} \gets$Find $l$ satisfying $|H_{p,l}| \geq |H_{p,q}|$ for all $q \in \mathcal{A}$
     \State $\Omega_p \gets \Omega_p \cup \{l^{*}\}$
     \State $\mathcal{A} \gets \mathcal{A} - \{l^{*}\}$
     \State Update $R_p$ according to (\ref{eq:cap})
      \EndFor
\While{$\mathcal{A} \neq \oslash$}
\State $p^{*}\gets $Find $p$ satisfying $R_p/\eta_p \leq R_q/\eta_q$ for all $q$ ($1\leq q \leq P$)
\State $l^{*} \gets$Find $l$ satisfying $|H_{p^{*},l}| \geq |H_{p^{*},q}|$ for all $q \in \mathcal{A}$
\State $\Omega_{p^{*}} \gets \Omega_{p^{*}} \cup \{l^{*}\}$
     \State $\mathcal{A} \gets \mathcal{A} - \{l^{*}\}$
     \State Update $R_{p^{*}}$ according to (\ref{eq:cap})
\EndWhile
\EndProcedure \\
\end{algorithmic}
\end{algorithm}

\subsection{The proposed video decoder}
\label{sec:decoder}
Each receiver first demodulates the received samples of each frame. The samples in each packet are then divided by the corresponding $g_{p,l}$. The samples of all blocks are then extracted from the received packets by performing the reverse operations of the packetization procedure. 
After obtaining the CS samples of all blocks in each frame, the BCS-SPL algorithm is applied on the obtained samples, to obtain an initial approximation of the transmitted frame. If the transmitted frame is a residual frame, then the last reconstructed frame is added to the initial approximation to get an approximation of the original P frame. Let $\hat{\mathbf{X}}$ be the obtained reconstructed frame. Note that since the encoder does not know the exact reconstruction frame at the decoder, this brings drifting errors. To further improve the reconstruction quality of $\hat{\mathbf{X}}$ and reduce drifting errors, we propose an improved version of the original BCS-SPL algorithm in which all steps are similar to the original BCS-SPL algorithm, however, with two main differences: 1) Unlike the original BCS-SPL algorithm that uses a fixed transform like DCT, the proposed algorithm uses an adaptive transform to obtain sparser transform coefficients. Note that the reconstruction quality of the CS reconstruction algorithms like BCS-SPL significantly depends on the sparsity of the transform coefficients \cite{fowler_conf}. Hence, it is reasonable to use adaptive transforms to achieve sparser transform coefficients. Also, adaptive transforms may represent signals better than fixed transforms \cite{dict1, dict2}. 2) To impose sparsity and reduce noise and distortions produced during the iterative reconstruction procedure, a hard-thresholding operator is used in the original BCS-SPL algorithm \cite{bcs_spl}. Instead of the hard-thresholding operator, we propose an adaptive soft-thresholding operator in the proposed reconstruction algorithm to achieve better reconstruction quality. 

In the sequel, we first present the proposed adaptive sparsifying transform in Section \ref{sec:transform}. We then propose the adaptive soft-thresholding operator in Section \ref{sec:thresholding}. The proposed frame reconstruction algorithm is described in Algorithm \ref{alg:recon}.

\subsubsection{The proposed adaptive transform}
\label{sec:transform}
Since adjacent frames are usually very similar and correlated, we can exploit this property to improve the reconstruction quality of the current frame. Here, we propose an adaptive sparsifying transform, which exploits this property.

Let $\hat{\mathbf{x}}_j$ be the vectortized $j$-th block in $\hat{\mathbf{X}}$ for $j=1,\cdots,M$. To build the proposed adaptive transform, for each block $\hat{\mathbf{x}}_j$ in the current frame, we first find the $K$ most similar blocks in the previously-reconstructed frame in a spatial window of size $S\times S$ pixels whose center location is the same as the center location of $\hat{\mathbf{x}}_j$. Here, we use the mean square error (MSE) as the block similarity metric. All the obtained $K$ blocks are then vectorized. Let $\mathbf{w}_j^k$ be the mean-centered and vectorized version of the $k$-th similar block for $\hat{\mathbf{x}}_j$. We then compute the covariance matrix $\mathbf{Z}_j$ of $\{\mathbf{w}_j^k\}_{k=1}^K$ as follows: 
\begin{equation}
\label{eq:Z}
\mathbf{Z}_j = \frac{1}{K} \sum_{k=1}^K \mathbf{w}_j^k (\mathbf{w}_j^k)^t,
\end{equation}
where $t$ denotes the transpose operator. The eigenvectors of $\mathbf{Z}_j$ are then computed to obtain $\mathbf{Z}_j= \mathbf{U}_j \mathbf{D}_j \mathbf{U}_j^t$, where $\mathbf{U}_j$ contains the eigenvectors, and $\mathbf{D}_j$ is a diagonal matrix that contains the corresponding eigenvalues of $\mathbf{Z}_j$ sorted in decreasing order. The obtained eigenvectors can be considered as the principal components of the most similar blocks to $\hat{\mathbf{x}}_j$. Hence, $\hat{\mathbf{x}}_j$ can be reconstructed using a small number of such components. Therefore, we propose to use $\pmb\Psi_j = \mathbf{U}_j$ as the adaptive sparsifying transform. Note that the proposed adaptive transform is implemented only at the decoder just to provide sparser transform coefficients for better frame reconstruction. But as a future work, we intend to change the structure so as to benefit from this transform at the encoder side as well. 

\subsubsection{The proposed soft-thresholding scheme}
\label{sec:thresholding}
As mentioned in Section \ref{sec:BCS}, in the BCS-SPL algorithm, to impose sparsity and reduce the blocking artifacts produced during the iterative procedure, a hard-thresholding operator is used. Since this is similar to a de-noising operator, its performance  has an important impact on the reconstruction quality. This operator sets to zero those transform coefficients whose amplitude is smaller than a threshold. Hence, accurate estimation of such thresholds is very important. Here, we propose an adaptive soft-thresholding operator to be used with the proposed adaptive transform in the BCS-SPL algorithm.  

Since $\pmb\Psi_j$ is computed from the blocks in the previously-reconstructed frame at the decoder, the transform coefficients of $\hat{\mathbf{x}}_j$, i.e. $\hat{\mathbf{v}}_j =\boldsymbol \Psi_j \hat{\mathbf{x}}_j$, are noisy due to the channel noise as well as the iterative BCS-SPL procedure used to reconstruct the previous and current frames. Although $\pmb\Psi_j$ is applied only at the decoder, we can postulate that a similar transform, $\pmb\Psi_j^o$, exists at the encoder, where by applying it on the noise-free $\mathbf{x}_j$, we can obtain ideal (noise-free) transform coefficients $\mathbf{v}_j =\boldsymbol \Psi_j^o \mathbf{x}_j$. Of course, $\pmb\Psi_j^o$ isn't actually applied in the encoder, but we simply model the noisy coefficients at the decoder as noise-contaminated versions of noise-free coefficients, which we consider as existing at the encoder:
$\hat{\mathbf{v}}_j = \mathbf{v}_j + \mathbf{n}_j,$ where $\mathbf{n}_j$ is the noise vector.

To reconstruct $\mathbf{x}_j$ from $\hat{\mathbf{x}}_j$, the added noise must be removed or reduced. In other words, the goal is to estimate (recover) noise-free (ideal) coefficients, $\mathbf{v}_j$, using the available noisy coefficients $\hat{\mathbf{v}}_j$. 
Since the noisy coefficients are obtained by $\pmb\Psi_j$, and this transform is essentially the PCA computed from some blocks in the previously-reconstructed frame, the transforms at the encoder and decoder are actually different. In other words, in general, $\pmb\Psi_j \neq \pmb\Psi_j^o$.
However, based on the results in \cite{matrix1}, if the largest eigenvalue of the covariance matrix at the encoder is not repeated, then the eigenvector corresponding to the largest eigenvalue of the covariance matrix at the decoder will be similar to that of the covariance matrix at the encoder. Also, from the random matrix theory \cite{random_matrix}, we know that repeated eigenvalues are rare, so we consider the case of the largest eigenvalue not being repeated as typical. Under these conditions, \cite{hanieh} shows that $E\{\pmb\Psi_j\} \approx \pmb\Psi_j^o$, where $E\{\cdot\}$ denotes the expectation operator. Therefore, we can assume that the expected transform at the decoder is approximately equal to the transform at the encoder. 

\begin{figure}
\centering
\includegraphics[scale=0.36]{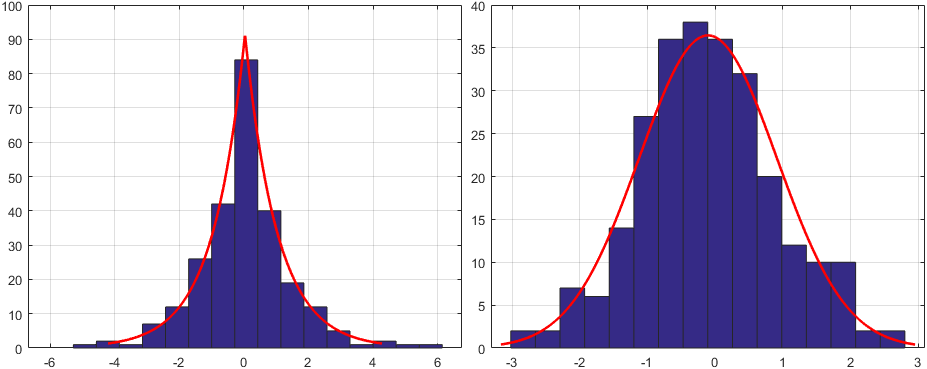}
\caption{Left: the empirical distribution of noise-free transform coefficients can be modeled by a Laplacian distribution (the red curve). Right: the empirical distribution of noise can be modeled by a Gaussian distribution (the red curve)}
\label{fig:dist}
\end{figure}

\begin{algorithm}
\caption{The proposed frame reconstruction algorithm.}
\label{alg:recon}
\begin{algorithmic}[1]
\Procedure{$\mathbf{X}=$recon\_frame}{$\mathcal{Y},\mathcal{P},k_{\max}, \epsilon$}
\State \textbf{Initialize:} Set $k=0$.
\State For each block $j$ do $\mathbf{x}_j^k=\pmb\Phi_j^t\mathbf{y}_j$ or apply the original BCS-SPL algorithm to obtain $\mathbf{x}_j^k$, and consequently $\mathbf{X}^k$.
    \Repeat
\State $\hat{\mathbf{X}}^k \gets$ SmoothingFilter($\mathbf{X}^k$) 
      \For{each block $j$}
        \State $\tilde{\mathbf{x}}_j^k \gets \hat{\mathbf{x}}_j^k + \pmb\Phi_j^t(\mathbf{y}_j-\pmb\Phi_j\hat{\mathbf{x}}_j^k)$
	\State Compute $\pmb\Psi_j$ using (\ref{eq:Z})
	\State $\hat{\mathbf{v}}_j^k \gets \pmb\Psi_j \tilde{\mathbf{x}}_j^k$
	\State $\nu \gets \text{median}(\{\hat{\mathbf{v}}_j^k\}|\forall j)$
	\State $\sigma_{n_j}^k \gets \text{median}(\{|\hat{\mathbf{v}}_j^k-\nu|\}|\forall j)/0.6745$
	\State $(\sigma_j^2)^k \gets  \max\Big(0,\frac{1}{\gamma} \sum_{j^{'}\in \mathcal{N}(j)}(\hat{\mathbf{v}}_{j^{'}}^k)^2-(\sigma_{n_j}^k)^2\Big)$
	\State $\mathbf{v}_j^k \gets \text{sgn}(\hat{\mathbf{v}}_j^k)\max\Big(0,|\hat{\mathbf{v}}_j^k|-\frac{\sqrt{2}(\sigma_{n_j}^k)^2}{\sigma_j^k}\Big)$
	\State $\bar{\mathbf{x}}_j^k \gets \pmb\Psi_j^{-1} \mathbf{v}_j^k$
	\State $\mathbf{x}_j^{k+1} \gets \bar{\mathbf{x}}_j^k + \pmb\Phi_j^t(\mathbf{y}_j-\pmb\Phi_j\bar{\mathbf{x}}_j^k)$
      \EndFor
	\State Compute error $e^{k+1} \gets \|\mathbf{X}^{k+1} - \tilde{\mathbf{X}}^{k}\|_2$
	\State $k \gets k+1$
	\Until{$|e^{k}-e^{k-1}|<\epsilon$ or $k=k_{\max}$}
	\State $\mathbf{X} \gets \mathbf{X}^{k}$
\EndProcedure \\

Note: $\mathcal{Y}=\{\mathbf{y}_1,\cdots,\mathbf{y}_M\}$ and $\mathcal{P}=\{\pmb\Phi_1,\cdots,\pmb\Phi_M\}$. $k_{\max}$ is the maximum number of iterations, and $\epsilon$ denotes the tolerance.
$\mathbf{x}_j^k$ denotes the $j$-th block of frame $\mathbf{X}$ in the $k$-th iteration.
\end{algorithmic}
\end{algorithm}

Hence, the estimation of $\mathbf{v}_j$ can be achieved using $\pmb\Psi_j$. For this purpose, Bayesian estimation methods like the Maximum a Posteriori (MAP) method can be used for which a prior knowledge about the noiseless coefficients $\mathbf{v}_j$ is needed.  Since the transform coefficients are usually sparse, based on our experiments, we found that the empirical distribution of $\mathbf{v}_j$ can be modeled by a Laplacian distribution as 
$\text{Pr}_{\mathbf{v}}(v_j)=\frac{1}{\sqrt{2}\sigma_j} \exp\Big(\frac{-\sqrt{2}|v_j|}{\sigma_j}\Big), $
where $\sigma_j$ is the standard deviation of the noiseless transform coefficients. In fact, in our experiments, we generated the histogram (statistical distribution) of noise-free transform coefficients of all $16\times 16$ blocks in the first 100 luminance frames of two standard video sequences (\textit{Bus} and \textit{Flower Garden}). We then computed the mean distribution of all the blocks. The obtained distribution is shown in the left image in Fig. \ref{fig:dist}. As seen from this image, such a distribution can be modeled by a Laplacian distribution.

In our expeirments, we also observed that the statistical distribution of $\mathbf{n}_j$ can be modeled by a Gaussian distribution as 
$\text{Pr}_{\mathbf{n}}(n_j)=\frac{1}{\sqrt{2}\sigma_{n_j}} \exp\Big(\frac{-n_j^2}{2\sigma_{n_j}^2}\Big), $
where $\sigma_{n_j}$ is the standard deviation of the noise, which can be estimated by various noise estimation methods. In this paper, we use the classic robust median estimator \cite{median} for this purpose, which estimates $\sigma_{n_j}$  as
$\sigma_{n_j} = \frac{\text{median}(\{|\hat{\mathbf{v}}_j-\nu|\})}{0.6745}, $
where $\nu = \text{median}(\{\hat{\mathbf{v}}_j\})$.
The right image in Fig. \ref{fig:dist} shows the mean distribution of  $\mathbf{n}_j$ of all $16\times 16$ blocks in the first 100 luminance frames of the two  sequences. It can be seen that such a distribution can be modeled by a Gaussian distribution.

To estimate $\sigma_j$, we use the Maximum Likelihood (ML) estimator on the noisy transform coefficients within a square neighborhood around $j$, i.e. $\mathcal{N}(j)$ as follows \cite{ML}:
\begin{multline}
\sigma_j^2 = \argmax_{\sigma_j^2 } \prod_{j^{'}\in \mathcal{N}(j)} \text{Pr}_{\hat{\mathbf{v}}|\sigma} \Big(\hat{\mathbf{v}}_{j^{'}}|\sigma_j^2\Big) = \\  \max\Big(0,\frac{1}{\gamma} \sum_{j^{'}\in \mathcal{N}(j)}\hat{\mathbf{v}}_{j^{'}}^2-\sigma_{n_j}^2\Big) ,
\end{multline}
where $\gamma$ is the number of elements in $\mathcal{N}(j)$, and $\text{Pr}_{\hat{\mathbf{v}}|\sigma}(\cdot)$ is a Gaussian distribution with zero mean and variance $\sigma_j^2 + \sigma_{n_j}^2$.

To estimate $\mathbf{v}_j$ using $\hat{\mathbf{v}}_j$, the MAP estimator is used as follows:
\begin{equation}
\mathbf{v}_j = \argmax_{\mathbf{v}_j} \Big\{ \text{Pr}_{\mathbf{v}|\hat{\mathbf{v}}} \big(\mathbf{v}_j|\hat{\mathbf{v}}_{j}\big)\Big\},
\end{equation}
where, using the Bayes' rule, we obtain:
\begin{multline}
\label{eq:bayes}
\mathbf{v}_j = \argmax_{\mathbf{v}_j} \Big\{ \text{Pr}_{\hat{\mathbf{v}}|\mathbf{v}} \big(\hat{\mathbf{v}}_{j}|\mathbf{v}_j\big) \text{Pr}_{\mathbf{v}}\big(\mathbf{v}_j\big)\Big\} = \\ \argmax_{\mathbf{v}_j} \Big\{ \text{Pr}_{\mathbf{n}}\big(\hat{\mathbf{v}}_j-\mathbf{v}_j\big) \text{Pr}_{\mathbf{v}}\big(\mathbf{v}_j\big) \Big\}.
\end{multline}
We can show that (\ref{eq:bayes}) is equivalent to the following equation:
\begin{equation}
\mathbf{v}_j = \argmin_{\mathbf{v}_j} \Big\{ \frac{1}{2}|\hat{\mathbf{v}}_j-\mathbf{v}_j|^2 + \frac{\sqrt{2}\sigma_{n_j}^2}{\sigma_j}|\mathbf{v}_j|\Big\},
\end{equation}
which yields the following solution:
\begin{equation}
\label{eq:soft}
\mathbf{v}_j = \text{sgn}(\hat{\mathbf{v}}_j)\max\Big(0,|\hat{\mathbf{v}}_j|-\frac{\sqrt{2}\sigma_{n_j}^2}{\sigma_j}\Big),
\end{equation}
where $\text{sgn}(\cdot)$ is the sign function. In fact, (\ref{eq:soft}) is the popular soft shrinkage function \cite{soft_thresholding}, which is defined as:
$\xi_\omega(y) = \text{sgn}(y)\max(0,|y|-\omega).$ Hence, we use (\ref{eq:soft}) as the soft thresholding operator in the proposed BCS reconstruction algorithm.

\begin{table}
\caption{Comparing the performance of various methods on \textit{Flower Garden} at four different target rates ($T_{tot}^1, \cdots, T_{tot}^4$) under three different CSNR levels ($\theta$).}
\label{tab:flower}
\centering
\scriptsize
\renewcommand{\tabcolsep}{0.02cm}
\begin{tabular}{|c|c|c|c|c|c|c|c|c|c|c|c|c|c|}
\hline
$\theta$&Method &\multicolumn{4}{c|}{\textit{PSNR}}&\multicolumn{4}{c|}{\textit{ST-RRED}}&\multicolumn{4}{c|}{\textit{MS-SSIM}}\\
\cline{3-14}
& &$T_{tot}^1$&$T_{tot}^2$&$T_{tot}^3$&$T_{tot}^4$&$T_{tot}^1$&$T_{tot}^2$&$T_{tot}^3$&$T_{tot}^4$&$T_{tot}^1$&$T_{tot}^2$&$T_{tot}^3$&$T_{tot}^4$\\ \cline{1-14}
&\cite{softcast1}&17.12&17.45&17.95&18.02&24.88&14.55&11.12&10.37&0.6411&0.6899&0.6974&0.7298 \\  \cline{2-14}
&\cite{dcs}&17.65&17.98&18.18&18.35&23.44&13.38&9.79&8.49&0.6498&0.6915&0.7153&0.7326  \\  \cline{2-14}
15&\cite{ardcs_cast}&17.75&18.92&19.11&19.85&10.11&3.88&2.92&2.43&0.7622&0.7854&0.7913&0.7929  \\  \cline{2-14}
&\cite{parcastplus}&18.32&19.41&19.98&20.04&8.56&2.50&2.41&2.05&0.7682&0.7902&0.8041&0.8098  \\  \cline{2-14}
&Proposed&\textbf{18.79}&\textbf{19.93}&\textbf{20.32}&\textbf{20.36}&\textbf{8.38}&\textbf{1.81}&\textbf{1.80}&\textbf{1.75}&\textbf{0.7795}&\textbf{0.8062}&\textbf{0.8147}&\textbf{0.8176} \\
\hhline{|=|=|=|=|=|=|=|=|=|=|=|=|=|=|}

&\cite{softcast1}&18.98&19.95&20.95&21.96&16.81&10.92&7.63&4.52&0.8289&0.8603&0.8814&0.8955 \\  \cline{2-14}
&\cite{dcs}&19.20&20.11&20.98&21.89&16.16&9.72&5.61&4.61&0.8301&0.8623&0.8817&0.8949  \\  \cline{2-14}
25&\cite{ardcs_cast}&21.05&24.82&26.15&27.19&9.65&3.58&2.87&2.33&0.8912&0.9215&0.9341&0.9369 \\  \cline{2-14}
&\cite{parcastplus}&21.92&25.12&26.95&27.84&8.21&2.45&2.01&1.45&0.8987&0.9291&0.9377&0.9410 \\  \cline{2-14}
&Proposed&\textbf{22.52}&\textbf{25.70}&\textbf{27.64}&\textbf{28.47}&\textbf{7.80}&\textbf{1.64}&\textbf{0.51}&\textbf{0.59}&\textbf{0.9081}&\textbf{0.9357}&\textbf{0.9463}&\textbf{0.9497}  \\
\hhline{|=|=|=|=|=|=|=|=|=|=|=|=|=|=|}

&\cite{softcast1}&19.21&20.19&21.20&22.21&15.99&9.09&6.23&3.78&0.8756&0.9100&0.9318&0.9492  \\  \cline{2-14}
&\cite{dcs}&19.51&20.55&21.61&22.77&15.65&8.91&5.49&3.46&0.8808&0.9160&0.9366&0.9511 \\  \cline{2-14}
35&\cite{ardcs_cast}&22.26&26.92&30.28&32.19&9.12&3.56&0.62&0.44&0.9251&0.9573&0.9668&0.9701  \\  \cline{2-14}
&\cite{parcastplus}&23.02&27.56&30.92&32.86&8.05&2.36&0.45&0.28&0.9289&0.9607&0.9710&0.9755  \\  \cline{2-14}
&Proposed&\textbf{23.62}&\textbf{28.16}&\textbf{31.44}&\textbf{33.41}&\textbf{7.13}&\textbf{1.59}&\textbf{0.26}&\textbf{0.11}&\textbf{0.9329}&\textbf{0.9623}&\textbf{0.9740}&\textbf{0.9792} \\
 \hline
\end{tabular}
\end{table}

\begin{table}
\caption{Comparing the performance of various methods on \textit{Bus} at four different target rates ($T_{tot}^1, \cdots, T_{tot}^4$) under three different CSNR levels ($\theta$).}
\label{tab:bus}
\centering
\scriptsize
\renewcommand{\tabcolsep}{0.02cm}
\begin{tabular}{|c|c|c|c|c|c|c|c|c|c|c|c|c|c|}
\hline
$\theta$&Method &\multicolumn{4}{c|}{\textit{PSNR}}&\multicolumn{4}{c|}{\textit{ST-RRED}}&\multicolumn{4}{c|}{\textit{MS-SSIM}}\\
\cline{3-14}
& &$T_{tot}^1$&$T_{tot}^2$&$T_{tot}^3$&$T_{tot}^4$&$T_{tot}^1$&$T_{tot}^2$&$T_{tot}^3$&$T_{tot}^4$&$T_{tot}^1$&$T_{tot}^2$&$T_{tot}^3$&$T_{tot}^4$\\ \cline{1-14}
&\cite{softcast1}&16.80&19.79&20.78&21.79&19.98&17.67&16.65&15.78&0.7735&0.8301&0.8668&0.8759  \\  \cline{2-14}
&\cite{dcs}&17.22&19.21&19.66&20.52&18.66&16.22&15.82&15.51&0.7782&0.8355&0.8632&0.8720 \\  \cline{2-14}
15&\cite{ardcs_cast}&18.92&20.06&20.89&21.26&14.52&12.25&9.29&2.82&0.7883&0.8412&0.8766&0.8869   \\  \cline{2-14}
&\cite{parcastplus}&19.21&20.45&21.28&22.01&12.95&9.28&6.22&2.99&0.7921&0.8489&0.8802&0.8952  \\  \cline{2-14}
&Proposed&\textbf{19.95}&\textbf{21.01}&\textbf{21.89}&\textbf{22.42}&\textbf{10.65}&\textbf{3.58}&\textbf{1.68}&\textbf{0.71}&\textbf{0.8071}&\textbf{0.8538}&\textbf{0.8841}&\textbf{0.9020 } \\
\hhline{|=|=|=|=|=|=|=|=|=|=|=|=|=|=|}

&\cite{softcast1}&20.71&21.71&22.70&23.70&17.89&16.83&15.66&15.42&0.8160&0.8600&0.8798&0.8909 \\  \cline{2-14}
&\cite{dcs}&20.52&22.00&23.28&24.69&16.53&15.75&15.52&15.30&0.8175&0.8647&0.8890&0.9061 \\  \cline{2-14}
25&\cite{ardcs_cast}&21.21&23.08&25.12&26.89&13.61&11.40&8.69&2.36&0.8256&0.8763&0.9068&0.9280  \\  \cline{2-14}
&\cite{parcastplus}&21.66&23.78&25.73&27.25&11.55&8.66&5.32&2.39&0.8305&0.8806&0.9116&0.9321\\  \cline{2-14}
&Proposed&\textbf{22.15}&\textbf{24.36}&\textbf{26.38}&\textbf{27.96}&\textbf{9.73}&\textbf{3.37}&\textbf{1.29}&\textbf{0.67}&\textbf{0.8358}&\textbf{0.8861}&\textbf{0.9193}&\textbf{0.9398} \\
\hhline{|=|=|=|=|=|=|=|=|=|=|=|=|=|=|}

&\cite{softcast1}&21.02&22.03&23.00&24.02&16.36&15.89&14.89&14.20&0.8165&0.8603&0.8806&0.8932  \\  \cline{2-14}
&\cite{dcs}&21.66&23.23&24.65&26.27&15.27&14.80&14.43&14.21&0.8178&0.8651&0.8899&0.9075 \\  \cline{2-14}
35&\cite{ardcs_cast}&21.42&24.34&26.83&28.76&12.52&9.63&6.89&2.33&0.8278&0.8826&0.9101&0.9321  \\  \cline{2-14}
&\cite{parcastplus}&21.95&24.82&27.15&29.23&10.55&7.55&4.50&1.80&0.8308&0.8868&0.9156&0.9359  \\  \cline{2-14}
&Proposed&\textbf{22.57}&\textbf{25.24}&\textbf{27.80}&\textbf{29.98}&\textbf{9.49}&\textbf{3.33}&\textbf{1.26}&\textbf{0.58}&\textbf{0.8397}&\textbf{0.8907}&\textbf{0.9246}&\textbf{0.9457 }
 \\
 \hline
\end{tabular}
\end{table}

\section{Experimental Results}
\label{sec:results}
To evaluate the performance of the proposed system, we performed a number of objective and subjective experiments under different conditions. Specifically, we compared the performance of the proposed system with four other approaches to demonstrate the superiority of the proposed system. These four approaches are SoftCast \cite{softcast1}, DCS-cast \cite{dcs}, ARDCS-cast \cite{ardcs_cast}, and ParCast+ \cite{parcastplus}, which were introduced in Section \ref{sec:intro}. For the experiments, we used the following four standard video sequences in standard CIF ($352\times 288$) format: \textit{Flower Garden}, \textit{Bus}, \textit{Mother \& Daughter}, and \textit{Mobile Calendar}. Only the luminance channel of the first 100 frames of each video were used in the experiments. Therefore, the source bandwidth is 6.08 MHz. The GOP size was set to 5. For computing the adaptive transform in (\ref{eq:Z}), we experimentally set $S=32$, and $K=10$. In fact, in our experiments, we found that larger values for $S$ and $K$ may not necessarily improve the performance of the proposed system. Hence, to keep the computational complexity reasonable, we chose these values. Also, for the proposed frame reconstruction algorithm, we used $3\times 3$ Wiener filter, and set $\epsilon=10^{-6}$ and $\gamma=9$. These values were found experimentally by trial and error. Similar to most of the traditional wireless communication systems, the Rayleigh fading channels are assumed in our experiments. 

\subsection{Objective Experiments}
For the objective performance evaluation of various methods, we used the following three objective quality metrics: PSNR (peak signal-to-noise ratio), MS-SSIM \cite{msssim}, and ST-RRED \cite{strred}. PSNR is a  widely-used non-perceptual full-reference image/video quality metric, that can be used to measure the fidelity of an image/video reconstruction process in a pixel-wise manner. In our experiments, we used the mean PSNR of all frames of a given reconstruted video as the quality score of that video. Note that the larger PSNR, the better quality and vice versa.

MS-SSIM is a well-lnown full-reference perceptual image quality metric \cite{msssim} that offers superior image quality prediction performance as compared to several existing and state-of-the-art image quality metrics \cite{gmad}. In our experiments, we computed MS-SSIM score for each frame of a reconstructed video. We then considered the mean MS-SSIM score of all frames as the final MS-SSIM score of that video. Note that the larger MS-SSIM, the better quality and vice versa.

ST-RRED \cite{strred} is a well-known and high-performance reduced-reference perceptual video quality metric.
The results reported in \cite{strred} showed that the ``1/576'' version of ST-RRED (that uses 1/576th of the available reference video information) achieves high performance on a wide variety of video quality databases as compared to several other popular video quality metrics like VMAF \cite{vmaf}, VQM \cite{vqm}, and MOVIE \cite{movie}. ST-RRED provides a quality score for each frame of a given video. In our experiments, we used the mean ST-RRED (1/576) score of all frames of a reconstructed video as the final ST-RRED quality score of that video. Note that the smaller ST-RRED, the better quality and vice versa.

For the simulations, we encoded each video with each approach at four different target rates ($[T_{tot}^1,T_{tot}^2,T_{tot}^3,T_{tot}^4]=[2,3,4,5]\times 10^6$) under three different channel SNR (CSNR) levels, i.e. 15 dB, 25 dB, and 35 dB. The actual rates of all approaches were matched with each other with 0.5\% error. The three quality metrics were then applied on the resultant reconstructed videos at each tested rate for each of the five approaches. This way, for each approach, we obtain a rate-distortion (RD) curve for each video under each tested CSNR level. The obtained results for all the four video sequences are shown in Table \ref{tab:flower} to Table \ref{tab:mobile}. As seen from Table \ref{tab:flower}, the proposed system outperforms the other four methods in terms of all quality metrics under all the three CSNR levels. The second best-performing method is ParCast+ whose performance is a little worse than the proposed system. After ParCast+, the best method is ARDCS-cast whose performance is close to ParCast+. Finally, the worst perfoming models are SoftCast and DCS-Cast, respectively. The performance gap between these two methods and the other three methods is significantly high. The reason is that the rate produced by SoftCast is generally high as compared to the CS-based methods. Also, unlike the proposed system and ARDCS-cast who transmit residual frames, DCS-Cast transmits each frame as an I frame. Note that similar to the proposed system, ParCast+ utilizes the source data redundancy by matching the more important source components with higher-gain channel components. Therefore, we believe the good performance of ParCast+ is due to this feature. Similar traits can be observed in Table \ref{tab:bus} to \ref{tab:mobile}.

We utilized the well-known BD-PSNR \cite{bd} metric to measure the average difference between two RD curves in terms of PSNR difference. In order to be able to compare the RD performance of various approaches, we considered the proposed system as our baseline, and computed the BD-PSNR of the other four methods with respect to the proposed system. Similarly, we computed BD-ST-RRED and BD-MS-SSIM of all the obtained RD curves. The results are shown in Table \ref{tab:bd} for three different CSNR levels. In this table, we also mentioned the average of BD-PSNR, BD-ST-RRED, and BD-MS-SSIM scores over all the tested video sequences for each CSNR level. From these results, we observe that, the BD-PSNR and BD-MS-SSIM of all the methods under all the three CSNR levels are negative, meaning that the performance of the proposed system is better than the other tested methods. Also, the BD-ST-RRED scores of all the methods are positive, which again indicates the superiority  of the proposed system over the other tested methods under the ST-RRED metric. Looking across all the results, we observe that the ranking of various methods under each tested CSNR level is: 1) the proposed system, 2) ParCast+, 3) ARDCS-cast, 4) DCS-Cast, and 5) SoftCast. Also, the performance gap between the proposed system and other methods generally increases as the CSNR level increases. We believe the reason is related to the proposed adaptive transform. Note that as the CSNR level increases, the frame reconstruction quality increases as well. Hence, when recovering a frame, better similar blocks in the previous frame are used to construct the adaptive transforms in the current frame, which in turn, improves the reconstruction quality of the current frame.

In Table \ref{tab:block}, we compared the performance of the proposed system when using different block sizes using the three different video quality metrics at a CNSR level of 15 dB, and at a total target rate of $2\times 10^6$. These results were obtained by averaging the metric values across the four video sequences. As seen from these results, using $8\times 8$ blocks achieves the best results, and that is why we used $8\times 8$ blocks in all of our experiments.

We also analyzed the performance of the proposed system when using different values for $(\alpha,\beta,\gamma)$ defined in (\ref{eq:pi}). Since these parameters can be any positive real numbers, we examined only five values $(0,0.25,0.5,0.75,1)$ for each of these three parameter. We then found that the best performance is achieved when using $\alpha=\beta=\gamma=1$. In Table \ref{tab:fuse}, we reported the metric results for some other combinations as some sample results.


\begin{table}
\caption{Comparing the performance of the proposed system with other methods based on BD-PSNR, BD-ST-RRED, and BD-MS-SSIM over three different CSNR levels ($\theta$).}
\label{tab:bd}
\centering
\scriptsize
\renewcommand{\tabcolsep}{0.1cm}
\renewcommand{\arraystretch}{0.9}
\begin{tabular}{|c|c|c|c|c|c|c|c|}
\hline
$\theta$&Metric&Method&\textit{Flower}&\textit{Bus}&\textit{Mother\&}&\textit{Mobile}& Avg. \\
&&&\textit{Garden}&&\textit{Daughter}&\textit{Calendar}& \\
\hline
\cline{1-8}
&&\cite{softcast1}&-2.02&-2.20&-0.93&-2.54&-1.92 \\
\cline{3-8}
&BD-PSNR&\cite{dcs}&-1.54&-2.37&-1.19&-1.66&-1.69 \\
\cline{3-8}
&&\cite{ardcs_cast}&-1.01&-1.02&-0.92&-0.94&-0.97 \\
\cline{3-8}
&&\cite{parcastplus}&-0.45&-0.65&-0.58&-0.60&-0.57 \\
\cline{2-8}

&&\cite{softcast1}&13.95&11.73&6.80&14.45&11.73 \\
\cline{3-8}
&BD-ST-RRED&\cite{dcs}&12.55&10.53&5.86&12.55&10.37 \\
\cline{3-8}
15&&\cite{ardcs_cast}&1.64&5.37&4.85&5.19& 4.26\\
\cline{3-8}
&&\cite{parcastplus}&0.37&3.41&3.64&2.80& 2.55\\
\cline{2-8}

&&\cite{softcast1}&-0.1258&-0.0284&-0.0374&-0.0305& -0.0555\\
\cline{3-8}
&BD-MS-SSIM&\cite{dcs}&-0.1182&-0.0254&-0.0445&-0.0251& -0.0533\\
\cline{3-8}
&&\cite{ardcs_cast}&-0.0196&-0.0155&-0.0067&-0.0130& -0.0137\\
\cline{3-8}
&&\cite{parcastplus}&-0.0120&-0.0104&-0.0018&-0.0063& -0.0076\\
\hline
\cline{1-8}

&&\cite{softcast1}&-4.75&-2.28&-1.74&-3.19& -2.99\\
\cline{3-8}
&BD-PSNR&\cite{dcs}&-4.60&-2.14&-1.14&-2.32&-2.55 \\
\cline{3-8}
&&\cite{ardcs_cast}&-1.31&-1.08&-0.80&-1.16& -1.08\\
\cline{3-8}
&&\cite{parcastplus}&-0.61&-0.55&-0.49&-0.71& -0.59\\
\cline{2-8}

&&\cite{softcast1}&8.38&10.84&9.60&15.02& 10.96\\
\cline{3-8}
&BD-ST-RRED&\cite{dcs}&7.48&9.81&7.47&13.07& 9.45\\
\cline{3-8}
25&&\cite{ardcs_cast}&1.93&5.16&4.01&5.54& 4.16\\
\cline{3-8}
&&\cite{parcastplus}&0.69&2.94&1.85&3.13& 2.15\\
\cline{2-8}

&&\cite{softcast1}&-0.0742&-0.0265&-0.0256&-0.0625& -0.0472\\
\cline{3-8}
&BD-MS-SSIM&\cite{dcs}&-0.0731&-0.0220&-0.0239&-0.0527& -0.0429\\
\cline{3-8}
&&\cite{ardcs_cast}&-0.0153&-0.0106&-0.0061&-0.0193& -0.0128\\
\cline{3-8}
&&\cite{parcastplus}&-0.0086&-0.0059&-0.0036&-0.0091& -0.0068\\
\hline
\cline{1-8}

&&\cite{softcast1}&-6.64&-2.77&-3.96&-3.92& -4.32\\
\cline{3-8}
&BD-PSNR&\cite{dcs}&-6.28&-1.72&-3.06&-3.28& -3.58\\
\cline{3-8}
&&\cite{ardcs_cast}&-1.29&-1.07&-0.94&-1.13& -1.10\\
\cline{3-8}
&&\cite{parcastplus}&-0.59&-0.59&-0.64&-0.75& -0.64\\
\cline{2-8}

&&\cite{softcast1}&7.70&9.73&10.85&14.24& 10.63\\
\cline{3-8}
&BD-ST-RRED&\cite{dcs}&7.34&8.82&5.64&13.20& 8.75\\
\cline{3-8}
35&&\cite{ardcs_cast}&1.62&4.05&3.12&4.86& 3.41\\
\cline{3-8}
&&\cite{parcastplus}&0.72&2.12&1.84&2.48& 1.79\\
\cline{2-8}

&&\cite{softcast1}&-0.0517&-0.0303&-0.0275&-0.0556& -0.0413\\
\cline{3-8}
&BD-MS-SSIM&\cite{dcs}&-0.0467&-0.0259&-0.0200&-0.0515& -0.0360\\
\cline{3-8}
&&\cite{ardcs_cast}&-0.0072&-0.0115&-0.0054&-0.0122& -0.0091\\
\cline{3-8}
&&\cite{parcastplus}&-0.0033&-0.0078&-0.0034&-0.0092& -0.0059\\
\hline
\cline{1-8}

\end{tabular}
\end{table}

\begin{table}
\caption{Comparing the performance of the proposed system (averaged over the four sequences) when using different block sizes based on PSNR, ST-RRED, and MS-SSIM at a CSNR level of 15 dB, and at a total target rate of $2\times 10^6$.}
\label{tab:block}
\centering
\small
\begin{tabular}{|c|c|c|c|}
\hline
Metric&$8\times 8$&$16\times 16$&$32\times 32$\\
\hline
PSNR&\textbf{19.96}&19.45&19.28 \\
\hline
ST-RRED&\textbf{10.33}&11.03& 11.64\\
\hline
MS-SSIM&\textbf{0.8106}&0.8097&0.8082 \\
\hline
\end{tabular}
\end{table}

\begin{table}
\caption{Comparing the performance of the proposed system (averaged over the four sequences) when using different values for $(\alpha,\beta,\gamma)$ based on PSNR, ST-RRED, and MS-SSIM at a CSNR level of 15 dB, and at a total target rate of $2\times 10^6$.}
\label{tab:fuse}
\centering
\small
\begin{tabular}{|c|c|c|c|}
\hline
$(\alpha,\beta,\gamma)$&PSNR&ST-RRED&MS-SSIM\\
\hline
 $(1,1,1)$&\textbf{19.96}&\textbf{10.33}&\textbf{0.8106} \\
\hline
 $(1,1,0)$&19.11&13.32&0.7965 \\
\hline
 $(1,1,0.5)$&19.35&10.78&0.8101 \\
\hline
 $(0.5,1,1)$&18.69&15.87&0.7906 \\
\hline
 $(1,0.5,1)$&19.05&11.63&0.8085 \\
\hline
 $(1,0.5,0.5)$&19.01&11.86&0.8098 \\
\hline
 $(0.5,1,0.5)$&18.56&15.96&0.7932 \\
\hline
 $(0.5,0.5,1)$&18.43&16.25&0.7895 \\
\hline
\end{tabular}
\end{table}

\subsection{Subjective Experiments}
Finally, we performed a subjective experiment to compare the perceptual quality of sequences encoded by the proposed system versus sequences encoded using other the 4 tested methods. Similar to \cite{ours1}, we utilized a Two Alternative Forced Choice (2AFC) method \cite{2afc} to compare subjective video quality. In 2AFC, the participant is asked to make a choice between two alternatives, in this case, the video encoded using the
proposed method vs. video encoded using one of the four tested methods. All the 4 CIF sequences were used in the experiment. All sequences were encoded by all the five methods at a rate of $2\times 10^6$, and the CSNR level was 25 dB. In each trial, participants were shown two videos, side by side, at the same vertical position separated by 1 cm horizontally on a midgray background. Each video pair was shown for 10 seconds. After this presentation, a mid-gray blank screen was shown for 5 seconds. During this period, participants were asked to indicate on an answer sheet, which of the two videos looks better (Left or Right). They were asked to answer either Left or Right for each video pair, regardless of how certain they were of their response. Participants did not know which video was produced by the proposed system and which one was produced by the other tested method. Randomly chosen half of the trials had the video produced by the proposed system on the left side of the screen and the other half on the right side, in order to counteract side bias in the responses. This gave a total of $4\cdot 2 \cdot 4 = 24$ trials.

The experiment was run in a quiet room with 15 participants (10 male, 5 female, aged between 18 and 28). All participants had normal or corrected-to-normal vision. A 22-inch LGl monitor with brightness 320 cd/m$^2$ and resolution $1920\times 1080$ pixels was used in our experiments. The brightness and
contrast of the monitor were set to 75\%. The illumination in the room was about 260 Lux. The distance between the monitor and the subjects was fixed
at 80 cm. Each participant was familiarized with the task before the start of the experiment via a short printed instruction sheet. The total length of the experiment for each participant was approximately 6 minutes.

The results are shown in Table \ref{tab:sub}, in which we show the number of responses that showed preference for the proposed system vs. any of the other four tested methods.To test for statistical significance, we used a two-sided $\chi^2$ test \cite{chi}, with the null hypothesis that there is no preference for either method, i.e., that the votes for each method come from distributions with the same mean. Under this hypothesis, the expected number of votes in each trial is 15 for each method, because each video pair was shown twice to each of the 15 participants. The $p$-value \cite{chi} of the test is also shown in the table. As a rule of thumb, the null hypothesis is rejected when $p < 0.05$. When this happens in Table \ref{tab:sub}, it means that the two methods under the comparison cannot be considered to have the same subjective quality, since one of them has obtained a statistically significantly higher number of votes, and therefore seems to have better quality.
Looking across all results in Table \ref{tab:sub}, we observe that the $p$-value of all comparisons is below 0.05, and the number of votes for the proposed system is always much greater than the other tested methods. This confirms that the proposed system offers higher subjective quality as compared to the other methods on all sequences. Fig. 1 in the supplementary document shows a visual example for comparing the perceptual quality of various methods.

\subsection{Computational Complexity}
We implemented the proposed system in Matlab R2017b, and we used the Matlab code of the other tested methods in our experiments. We measured the average time (in msec) for encoding and decoding of a CIF video sequence with 100 frames of various methods on an Intel Core i7 CPU @3.36 GHz with 16 GB RAM. The results are shown in Table \ref{tab:time}. As seen from these results, the slowest method for both encoding and decoding among the tested methods is the proposed system, and the fastest method for encoding is DCS-Cast, and for decoding is SoftCast. As a future work, we intend to optimize the proposed system for speed.

\begin{table}
\caption{Comparing the performance of various methods on \textit{Mother \& Daughter} at four different target rates ($T_{tot}^1, \cdots, T_{tot}^4$) under three different CSNR levels ($\theta$).}
\label{tab:mother}
\centering
\scriptsize
\renewcommand{\tabcolsep}{0.02cm}
\begin{tabular}{|c|c|c|c|c|c|c|c|c|c|c|c|c|c|}
\hline
$\theta$&Method &\multicolumn{4}{c|}{\textit{PSNR}}&\multicolumn{4}{c|}{\textit{ST-RRED}}&\multicolumn{4}{c|}{\textit{MS-SSIM}}\\
\cline{3-14}
& &$T_{tot}^1$&$T_{tot}^2$&$T_{tot}^3$&$T_{tot}^4$&$T_{tot}^1$&$T_{tot}^2$&$T_{tot}^3$&$T_{tot}^4$&$T_{tot}^1$&$T_{tot}^2$&$T_{tot}^3$&$T_{tot}^4$\\ \cline{1-14}
&\cite{softcast1}&21.32&22.30&23.31&24.32&24.23&14.52&13.66&12.75&0.7798&0.7816&0.7846&0.7990 \\  \cline{2-14}
&\cite{dcs}&22.22&21.75&21.15&20.62&24.00&13.71&10.91&9.76&0.7681&0.7821&0.7856&0.7869 \\  \cline{2-14}
15&\cite{ardcs_cast}&21.82&22.18&22.42&23.05&20.66&16.85&11.33&9.61&0.8108&0.8125&0.8142&0.8286   \\  \cline{2-14}
&\cite{parcastplus}&22.11&22.56&22.90&23.32&19.60&14.92&10.58&8.70&0.8162&0.8176&0.8189&0.8301  \\  \cline{2-14}
&Proposed&\textbf{22.73}&\textbf{23.10}&\textbf{23.41}&\textbf{23.89}&\textbf{14.52}&\textbf{11.90}&\textbf{9.26}&\textbf{8.72}&\textbf{0.8175}&\textbf{0.8193}&\textbf{0.8206}&\textbf{0.8356 } \\
\hhline{|=|=|=|=|=|=|=|=|=|=|=|=|=|=|}

&\cite{softcast1}&29.13&30.13&31.13&32.12&13.52&12.50&12.15&11.95&0.9252&0.9335&0.9425&0.9455 \\  \cline{2-14}
&\cite{dcs}&30.12&30.83&31.05&31.06&11.60&10.16&10.08&8.98&0.9267&0.9378&0.9423&0.9447  \\  \cline{2-14}
25&\cite{ardcs_cast}&30.69&30.88&31.08&31.23&8.45&6.90&5.52&4.69&0.9497&0.9512&0.9525&0.9542  \\  \cline{2-14}
&\cite{parcastplus}&31.02&31.12&31.32&31.66&6.12&4.63&4.21&2.65&0.9515&0.9543&0.9563&0.9570\\  \cline{2-14}
&Proposed&\textbf{31.53}&\textbf{31.66}&\textbf{31.75}&\textbf{32.02}&\textbf{4.09}&\textbf{2.73}&\textbf{2.49}&\textbf{1.80}&\textbf{0.9564}&\textbf{0.9571}&\textbf{0.9574}&\textbf{0.9588}  \\
\hhline{|=|=|=|=|=|=|=|=|=|=|=|=|=|=|}

&\cite{softcast1}&32.96&33.95&34.93&35.95&13.50&12.05&11.65&11.10&0.9496&0.9589&0.9668&0.9702   \\  \cline{2-14}
&\cite{dcs}&33.30&35.22&36.72&37.99&7.67&7.37&7.19&7.13&0.9560&0.9686&0.9748&0.9785 \\  \cline{2-14}
35&\cite{ardcs_cast}&36.42&36.83&37.25&37.69&5.90&4.65&3.65&2.05&0.9762&0.9787&0.9800&0.9846   \\  \cline{2-14}
&\cite{parcastplus}&36.72&37.12&37.60&38.02&4.52&3.20&2.30&1.95&0.9789&0.9802&0.9811&0.9854 \\  \cline{2-14}
&Proposed&\textbf{37.37}&\textbf{37.80}&\textbf{38.17}&\textbf{38.54}&\textbf{2.34}&\textbf{1.49}&\textbf{1.09}&\textbf{0.90}&\textbf{0.9816}&\textbf{0.9844}&\textbf{0.9864}&\textbf{0.9880}   \\
 \hline
\end{tabular}
\end{table}

\begin{table}
\caption{Comparing the performance of various methods on \textit{Mobile Calendar} at four different target rates ($T_{tot}^1, \cdots, T_{tot}^4$) under three different CSNR levels ($\theta$).}
\label{tab:mobile}
\centering
\scriptsize
\renewcommand{\tabcolsep}{0.02cm}
\begin{tabular}{|c|c|c|c|c|c|c|c|c|c|c|c|c|c|}
\hline
$\theta$&Method &\multicolumn{4}{c|}{\textit{PSNR}}&\multicolumn{4}{c|}{\textit{ST-RRED}}&\multicolumn{4}{c|}{\textit{MS-SSIM}}\\
\cline{3-14}
& &$T_{tot}^1$&$T_{tot}^2$&$T_{tot}^3$&$T_{tot}^4$&$T_{tot}^1$&$T_{tot}^2$&$T_{tot}^3$&$T_{tot}^4$&$T_{tot}^1$&$T_{tot}^2$&$T_{tot}^3$&$T_{tot}^4$\\ \cline{1-14}
&\cite{softcast1}&15.33&16.33&17.32&18.35&22.09&19.22&17.87&17.03&0.8103&0.8288&0.8511&0.8652 \\  \cline{2-14}
&\cite{dcs}&16.56&17.05&17.87&18.06&19.66&17.55&16.73&16.56&0.8185&0.8305&0.8557&0.8645  \\  \cline{2-14}
15&\cite{ardcs_cast}&17.40&17.75&18.21&18.62&12.52&11.23&8.05&6.98&0.8290&0.8478&0.8655&0.8755   \\  \cline{2-14}
&\cite{parcastplus}&17.72&18.12&18.64&18.90&11.12&6.85&5.25&4.60&0.8322&0.8589&0.8760&0.8863  \\  \cline{2-14}
&Proposed&\textbf{18.38}&\textbf{18.71}&\textbf{19.06}&\textbf{19.40}&\textbf{7.77}&\textbf{4.78}&\textbf{3.17}&\textbf{2.10}&\textbf{0.8383}&\textbf{0.8655}&\textbf{0.8820}&\textbf{0.8935}  \\
\hhline{|=|=|=|=|=|=|=|=|=|=|=|=|=|=|}

&\cite{softcast1}&18.50&19.51&20.52&21.50&21.50&18.32&17.63&16.88&0.8155&0.8765&0.9088&0.9256 \\  \cline{2-14}
&\cite{dcs}&19.18&20.46&21.66&22.88&19.12&16.52&16.44&16.12&0.8282&0.8828&0.9134&0.9348 \\  \cline{2-14}
25&\cite{ardcs_cast}&20.10&21.89&23.22&24.11&12.05&10.99&6.61&3.75&0.8750&0.9088&0.9260&0.9380 \\  \cline{2-14}
&\cite{parcastplus}&20.56&22.11&23.89&24.78&10.55&6.55&3.23&2.78&0.8875&0.9155&0.9326&0.9488\\  \cline{2-14}
&Proposed&\textbf{21.37}&\textbf{22.91}&\textbf{24.18}&\textbf{25.32}&\textbf{6.54}&\textbf{3.80}&\textbf{1.95}&\textbf{1.19}&\textbf{0.8974}&\textbf{0.9241}&\textbf{0.9417}&\textbf{0.9543} \\
\hhline{|=|=|=|=|=|=|=|=|=|=|=|=|=|=|}

&\cite{softcast1}&19.02&20.02&21.03&22.04&19.65&18.22&17.05&16.23&0.8352&0.8905&0.9215&0.9432    \\  \cline{2-14}
&\cite{dcs}&19.37&20.75&22.12&23.59&18.89&16.29&16.06&15.85&0.8393&0.8941&0.9258&0.9477  \\  \cline{2-14}
35&\cite{ardcs_cast}&21.02&23.11&24.97&27.15&11.58&8.22&5.69&3.55&0.8952&0.9220&0.9431&0.9500    \\  \cline{2-14}
&\cite{parcastplus}&21.23&23.65&25.69&27.68&9.11&5.60&3.12&2.68&0.8965&0.9269&0.9488&0.9535\\  \cline{2-14}
&Proposed&\textbf{22.04}&\textbf{24.29}&\textbf{26.41}&\textbf{28.38}&\textbf{6.03}&\textbf{3.51}&\textbf{1.80}&\textbf{1.04}&\textbf{0.9073}&\textbf{0.9345}&\textbf{0.9526}&\textbf{0.9657}  \\
 \hline
\end{tabular}
\end{table}

\begin{table}
\caption{Subjective comparison of the proposed system against other methods}
\label{tab:sub}
\centering
\scriptsize
\begin{tabular}{|c|c|c|c|c|}
\hline
Method&\textit{Flower}&\textit{Bus}&\textit{Mother\&}&\textit{Mobile} \\
& \textit{Garden}& & \textit{Daughter} & \textit{Calendar} \\
\hline
\cline{1-5}
Proposed&26& 22 & 23&25  \\
\hline
\cite{parcastplus}&4& 8 & 7& 5 \\
\hline
$p$-value& 0.0001&0.0106 & 0.0105& 0.0003 \\
\hline
\cline{1-5}
Proposed& 22& 26& 27& 26  \\
\hline
\cite{ardcs_cast}&8 & 4& 3& 4 \\
\hline
$p$-value&0.0106 & 0.0001& 0.0001&  0.0001\\
\hline
\cline{1-5}
Proposed& 30& 30& 30&29  \\
\hline
\cite{dcs}& 0& 0& 0& 1 \\
\hline
$p$-value& $<10^{-4}$& $<10^{-4}$& $<10^{-4}$&$<10^{-4}$  \\
\hline
\cline{1-5}
Proposed& 30& 30& 30&30  \\
\hline
\cite{softcast1}& 0& 0& 0& 0 \\
\hline
$p$-value& $<10^{-4}$& $<10^{-4}$& $<10^{-4}$&$<10^{-4}$  \\
\hline
\cline{1-5}

\end{tabular}
\end{table}

\begin{table}
\caption{Average encoding/decoding time (msec) of various methods}
\label{tab:time}
\centering
\small
\renewcommand{\tabcolsep}{0.1cm}
\renewcommand{\arraystretch}{0.99}
\begin{tabular}{|c|c|c|c|c|c|}
\hline
&Proposed&\cite{parcastplus}&\cite{ardcs_cast}&\cite{dcs}&\cite{softcast1} \\
\hline
Encoding&6405&5630&4736&527&2308 \\ 
\hline
Decoding&27563&12680&16117&11535&2405 \\  \hline
\end{tabular}
\end{table}

\section{Conclusions}
\label{sec:conclusions}
In this paper, we presented a pseudo-analog soft video multicasting system using adaptive block-based compressed sensing. At the transmitter side of the proposed system, a target rate is first assigned to each frame of a given input video sequence according to the frames complexity, and a given total target rate. After that, different blocks in each frame are sampled adaptively according to their texture complexity and visual importance based on the frame's target rate. The generated samples are then packetized in an error-resilient manner, and the obtained packets are transmitted in a pseudo-analog manner through an OFDM channel after subchannel and power allocation. At the receiver side, the proposed BCS-based frame reconstruction algorithm is applied on the obtained samples to recover the transmitted frames. The reconstruction algorithm utilizes an adaptive sparsifying transform that exploits the similarity of adjacent frames as well as an adaptive soft-thresholding operator to improve the sparsity of the transform coefficients, thereby improving the frame reconstruction quality. Experimental results on various standard video sequences indicated that the proposed system has a higher performance as compared to several existing methods at different channel SNRs and different rates. As a future work, we intend to develop the hybrid version of the proposed system in which the base layer is encoded by a digital encoder and the enhancement layer is encoded by the proposed pseudo-analog system.

\bibliographystyle{IEEEtran}
\bibliography{ref}

\begin{IEEEbiography}[{\includegraphics[width=1in,height=1.25in,clip,keepaspectratio]{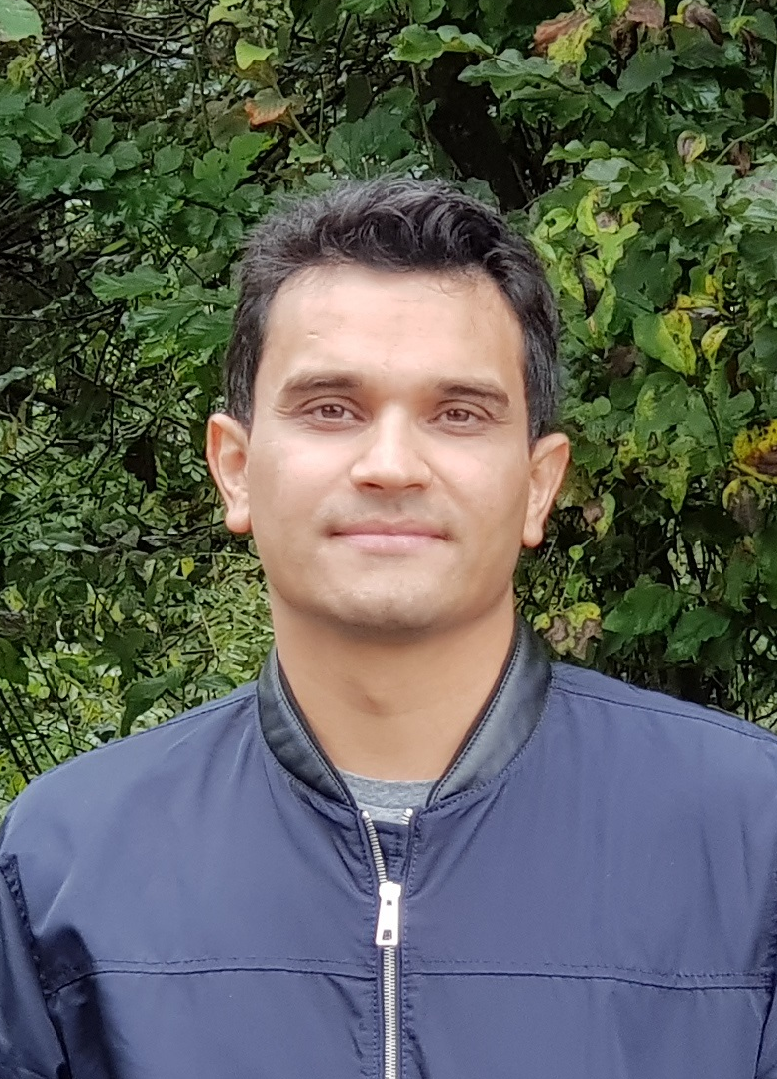}}]{Hadi Hadizadeh}
received the B.Sc.Eng. degree
in electronic engineering from the Shahrood University
of Technology, Shahrood, Iran, in 2005,
the M.S. degree in electrical engineering from the
Iran University of Science and Technology, Tehran,
Iran, in 2008, and the Ph.D. degree in engineering
science from Simon Fraser University, Burnaby,
BC, Canada, in 2013. He is currently an Associate
Professor with the Quchan University of Technology,
Quchan, Iran. His current research interests
include perceptual image/video coding, visual attention
modeling, error resilient video transmission, image/video processing,
computer vision, multimedia communication, and machine learning. He was
a recipient of the Best Paper Runner-up Award at ICME 2012 in Melbourne,
Australia, and the Microsoft Research and Canon Information Systems
Research Australia Student Travel Grant for ICME 2012. In 2013, he was
serving as the Vice Chair for the Vancouver Chapter of the IEEE Signal
Processing Society.
\end{IEEEbiography}

\begin{IEEEbiography}[{\includegraphics[width=1in,height=1.25in,clip,keepaspectratio]{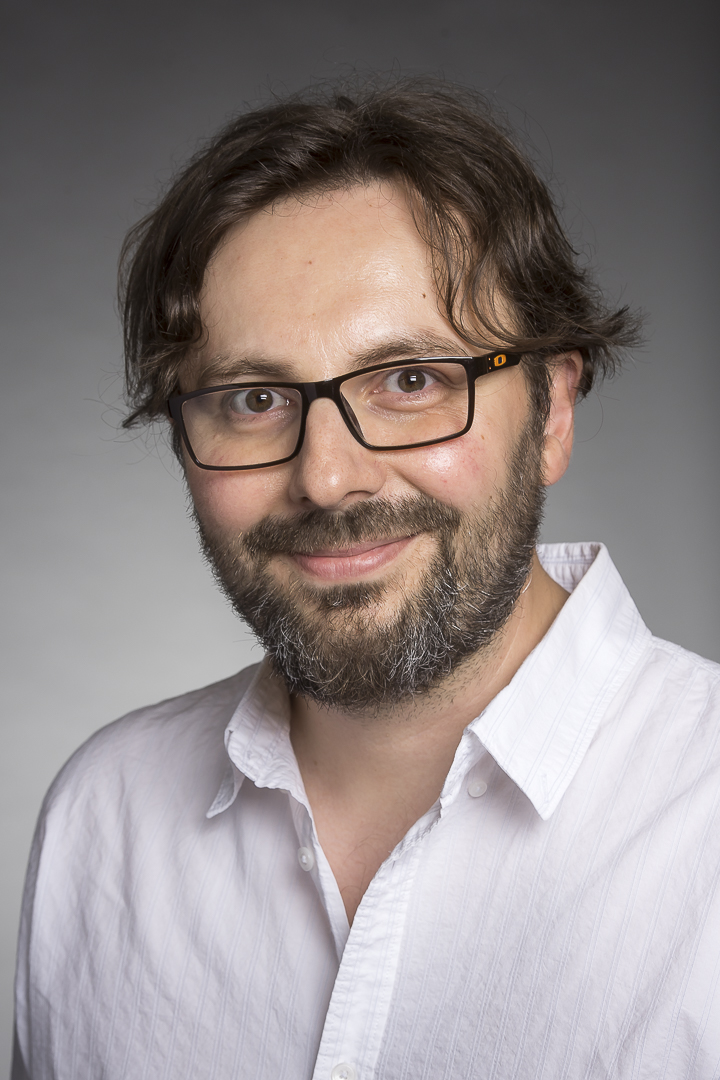}}]{Ivan V. Baji\'{c}}
 (S'99-M'04-SM'11) received the Ph.D. degree in Electrical Engineering from Rensselaer Polytechnic Institute, Troy, NY, in 2003. He is a Professor of Engineering Science and co-director of the Multimedia Lab at Simon Fraser University, Burnaby, BC, Canada. His research interests include signal processing and machine learning with applications to multimedia processing, compression, and collaborative intelligence. He has authored about a dozen and co-authored another ten dozen publications in these fields. Several of his papers have received awards, most recently at ICIP 2019. 

He was the Chair of the Media Streaming Interest Group of the IEEE Multimedia Communications Technical Committee from 2010 to 2012, and is currently an elected member of the IEEE Multimedia Signal Processing Technical Committee and the IEEE Multimedia Systems and Applications Technical Committee. He has served on the organizing and/or program committees of the main conferences in the field, and has received five revieweing awards, most recently at ICASSP 2019. He was an Associate Editor of \textsc{IEEE Transactions on Multimedia} and \textsc{IEEE Signal Processing Magazine}, and is currently an Area Editor of \textsc{Signal Processing: Image Communication}. He was the Chair of the IEEE Signal Processing Society Vancouver Chapter for several years, during which the Chapter received the Chapter of the Year Award from IEEE SPS. 
\end{IEEEbiography}

\end{document}